\documentclass[10pt,journal,twocolumn]{IEEEtran}
\newif\ifCLASSOPTIONromanappendices \CLASSOPTIONromanappendicestrue
\usepackage[utf8x]{inputenc}
\usepackage{a0size}         
\usepackage{amssymb}
\usepackage{amsmath}
\usepackage{multicol}
\usepackage[english]{babel}    
\usepackage{epsfig} 
\usepackage{bm}
\usepackage{bbm}
\usepackage{amsfonts,color,amsthm,amsmath, lscape,graphics}
\usepackage{url}
\usepackage{comment}
\usepackage{algorithm} 
\usepackage{algpseudocode}

\usepackage[font=footnotesize]{caption}
\usepackage[font=footnotesize]{subcaption}
\usepackage{subcaption}
\usepackage{epstopdf}
\usepackage{algorithm} 
\usepackage{algpseudocode}
\usepackage{cite}

\definecolor{awesome}{rgb}{1.0, 0.13, 0.32}
\usepackage{fix2col}

\addto\captionsenglish{}
\newcommand{\changef}[1]{{\color{black}#1}}

\usepackage{graphicx}  
\theoremstyle{plain}

\usepackage{multirow} 
\usepackage{relsize}  
\usepackage{bm}   
\usepackage{pifont}

\newcommand{\argmax}{\arg\!\max}

\begin{document}
\title{Massive Unsourced Random Access Based on  Uncoupled Compressive Sensing: Another Blessing of Massive MIMO}

\author{Volodymyr Shyianov, Faouzi~Bellili, \IEEEmembership{Member, IEEE}, Amine Mezghani, \IEEEmembership{Member, IEEE}, and Ekram Hossain, \IEEEmembership{Fellow, IEEE}
 \vspace{0.3cm}
\\\small E2-390 E.I.T.C,  75 Chancellor's Circle  Winnipeg, MB, Canada, R3T 5V6.
  \vspace{0.1cm}
  \\\small Emails:  shyianov@myumanitoba.ca, \{Faouzi.Bellili, Amine.Mezghani, Ekram.Hossain\}@umanitoba.ca.
  \vspace{0.3cm}
\thanks{The authors are with the Department of Electrical and Computer Engineering at the University of Manitoba, Winnipeg, MB, Canada.  This work was supported by the Discovery Grants Program of the Natural Sciences and Engineering Research Council of Canada (NSERC). }}

\maketitle
\begin{abstract}
We put forward a new algorithmic solution to the massive unsourced random access (URA) problem, by leveraging the rich spatial dimensionality offered by large-scale antenna arrays. This paper makes an observation that spatial signature is key to URA in massive connectivity setups. The proposed scheme relies on a slotted transmission framework but eliminates the need for  concatenated coding that was introduced in the context of the coupled compressive sensing (CCS) paradigm. Indeed, all existing works on CCS-based URA rely on an inner/outer tree-based encoder/decoder to stitch the slot-wise recovered sequences. This paper takes a different path by harnessing the nature-provided correlations between the slot-wise reconstructed channels of each user in order to put together its decoded sequences. The required slot-wise channel estimates and decoded sequences are first obtained through the hybrid generalized approximate message passing (HyGAMP) algorithm which systematically accommodates the multiantenna-induced group sparsity. Then, a channel correlation-aware clustering framework based on the expectation-maximization (EM) concept is used together with the Hungarian algorithm to find the slot-wise  optimal assignment matrices by enforcing two clustering constraints that are very specific to the problem at hand. Stitching is then accomplished by associating the decoded sequences to their respective users according to the ensuing assignment matrices. Exhaustive computer simulations reveal that the proposed scheme can bring performance improvements,  at high spectral efficiencies, as compared to  a state-of-the-art technique that investigates the use of large-scale antenna arrays in the context of massive URA.  
\end{abstract}

\begin{IEEEkeywords}
 Unsourced random access, massive connectivity, massive MIMO, channel estimation, clustering, hybrid approximate message passing.
\end{IEEEkeywords}


\section{Introduction}

\subsection{Background and Motivation}
\IEEEPARstart{M}{assive} random access in which a base station (BS) equipped 
with a large number of antennas is serving a large number of contending users has recently attracted considerable attention. This surge of interest is fueled by the need to satisfy the demand in wireless connectivity for many envisioned IoT applications such as massive machine-type communication (mMTC).  MTC has two distinct features \cite{dutkiewicz2017massive} that make them drastically different from human-type communications (HTC) around which previous cellular systems have mainly evolved: $i)$ machine-type devices (MTDs) require sporadic access to the network and $ii)$ MTDs usually transmit small data payloads using short-packet signaling. The sporadic access leads to the overall mMTC traffic being generated by an unknown and random subset of active MTDs (at any given transmission instant or frame). This calls for the development of scalable random  access protocols that are able to accommodate a massive number of MTDs. Short-packet transmissions, however, make the traditional grant-based access (with the associated scheduling overhead) fall short in terms of spectrum efficiency and latency, which are two key performance metrics in next-generation wireless networks. Hence, a number of grant-free random access schemes have been recently investigated within the specific context of massive connectivity (see \cite{Wei_Yu_paper} and references therein).    
In \textit{sourced}\footnote{As opposed to the unsourced case, sourced multiple access refers to the case where the BS is interested in both the messages and the identities of the users that generated them.} random access, grant-free transmissions often require two phases: $i)$ pilot sequences are first used to detect the active users and estimate their channels, then $ii)$ the identified active users are scheduled to transmit their messages.\\
In this context, it was shown that the joint device activity detection and channel estimation task can be cast as a compressed sensing (CS) problem; more precisely a useful CS variant called the multiple-measurement vector (MMV) in presence of multiple receive antennas.
 Among a plethora of CS recovery techniques, the approximate message passing (AMP) algorithm \cite{AMP} has attracted considerable attention within the framework of massive random access mainly due to the existence of simple scalar equations that track its dynamics, as rigorously analyzed in \cite{state_evolution}.\\ 
\indent Besides CS-based schemes, there is another line of work that has investigated the use of random access strategies based on conventional ALOHA \cite{ALOHA} and coded slotted ALOHA \cite{coded_access}. In many applications, however, the BS is interested in the transmitted messages only and not the IDs of the users, thereby leading to the so-called unsourced random access (URA). The information-theoretic work in \cite{5G_1} introduced a random coding existence bound for URA using a random Gaussian codebook with maximum likelihood-decoding at the BS. 
Moreover, popular multiple access schemes, e.g., ALOHA, coded slotted ALOHA, and treating interference as noise (TIN) were compared against the established fundamental limit, showing that none of them achieves the optimal predicted performance. The difficulty in achieving the underlying bound stems from the exponential (in blocklength) size of the codebooks  analyzed in \cite{5G_1}. \\\indent Recent works on URA have focused more on the algorithmic aspect of the problem by relying on the CS-based encoding/decoding paradigm in conjunction with a slotted transmission framework. The use of slotted transmissions is driven by the need to alleviate the inherent prohibitive computational burden of the underlying index coding problem. More specifically, in the coded/coupled compressive sensing (CCS) scheme \cite{5G_3}, the binary message of each user is partitioned into multiple  information bit sequences (or chunks). Then  binary linear block coding is used to couple the different sequences before using a random Gaussian codebook for actual transmissions over multiple slots. At the receiver side, inner CS-based decoding is first performed to recover the slot-wise transmitted sequences up to an unknown permutation. An outer tree-based decoder is then used to stitch the decoded binary sequences across different slots. A computationally tractable URA scheme has been recently proposed in \cite{fengler2019sparcs} --- based on the CCS framework --- wherein the authors exploit the concept of sparse regression codes (SPARCs) \cite{joseph2012least} to reduce the size of the required codebook matrix. The main idea of SPARCs is to encode information in structured linear combinations of the columns of a fixed codebook matrix so as to design a polynomial-time complexity encoder over the field of real numbers. In \cite{fengler2019sparcs}, the AMP algorithm was used as inner CS decoder and the state-evolution framework was utilized to analyze the performance of the resulting URA scheme.
Further extensions of the CSS framework for URA were also made in \cite{5G_4} where a low-complexity algorithm based on chirps for CS decoding was introduced. A number of other algorithmic solutions to the unsourced random access problem were also reported in  \cite{pradhan2019polar,amalladinne2019enhanced,pradhan2019joint}.  However, all the aforementioned works assume a single receive antenna  at the BS and it was only recently that the use of large-scale antenna arrays in the context of URA has been investigated in \cite{5G_2}. There, the authors use a low-complexity covariance-based CS (CB-CS) recovery algorithm \cite{Covariance} for activity detection, which iteratively finds the large-scale fading coefficients of all the users. Within the specific context of massive connectivity, CB-CS has the best known scaling law in terms of the required number of observations versus the number of active users. Above the CS regime \cite{fengler2019non}, i.e., more active users than observations, CB-CS algorithm generally requires a much smaller number of antennas than AMP-MMV to achieve the same level of performance. In the context of massive URA this particular regime of operation is most desirable since an  increase in the number of active users at a fixed number of observations leads to higher spectral efficiency. In \textit{sourced} random access, existing CS-based algorithms that reconstruct the entire channel matrix require large-size pilot sequences (i.e., a prohibitively large overhead) in presence of a large number of active users. To sidestep this problem, CB-SC relies rather on the use of receive antennas to identify the more active users by estimating their large-scale coefficients only. In the URA scenario, however, no user identification is required and the entire transmission frame is dedicated to data communication.

\subsection{Contributions}

We devise in this paper an algorithmic solution to the URA problem that can accommodate much more active users than the number of receive antenna elements at the BS. Assuming the channels to remain almost unchanged over the entire transmission period, the proposed scheme exploits the spatial channel statistics to stitch the decoded binary sequences among different slots thereby eliminating the need for concatenated coding as was done in all existing works on CCS-based URA \cite{5G_2,5G_3,5G_4}. In fact, the strong correlation between the slot-wise reconstructed channel vectors  pertaining to each active device already provides sufficient information for stitching its decoded sequences across the different slots. It is the task of the inner CS-based decoder to recover the support of the unknown  sparse vector and to estimate the users' channels in each slot. Each recovered support is used to decode the associated information bit sequences that were transmitted by all the  active users. Then, by clustering together the slot-wise reconstructed channels of each user, it will be possible to cluster/stitch its decoded sequences in order to recover its entire packet.\\
\indent Our CS-based decoder is based on a recent CS technique called the HyGAMP algorithm, which is able to account for the group sparsity in the underlying MMV model by incorporating latent Bernoulli random variables. HyGAMP runs loopy belief propagation coupled with Gaussian and quadratic approximations, for the propagated messages, which become increasingly accurate in the large system limits (i.e., large codebook sizes). At convergence, HyGAMP provides MMSE and MAP estimates of the users' channels and their activity-indicator Bernoulli random variables. It should be noted that HyGAMP is one of the varieties of CS algorithms that can be used in conjunction with clustering-based stitching. For instance, AMP-MMV \cite{ziniel2012efficient}, CoSaMP \cite{needell2009cosamp}, Group Lasso \cite{friedman2010note} or even any support recovery algorithm followed by least-squares channel estimation (e.g., \cite{fengler2019non, 5G_2}) can  all be envisaged. While the performance would vary depending on the particular choice of CS algorithm, those and other alternatives were not further explored in this work.
\\
\indent We further resort to the Gaussian-mixture expectation-maximization principle for channel clustering in combination with an integer optimization framework to embed two clustering constraints that are very specific to our problem. It will be seen that the newly proposed algorithm outperforms the state-of-the-art related techniques. In particular, our algorithm makes it possible to accommodate a larger total spectral efficiency with reasonable antenna array sizes while bringing in performance advantages in terms of the decoding error probability.  
    
\subsection{Organization of the Paper and Notations}
We structure the rest of this paper as follows. In Section II, we introduce the system model.
 In Section III, we describe the HyGAMP-based inner CS encoder/decoder, as well as, the clustering-based stitching procedure of the decoded sequences. In Section IV, we assess the performance of the proposed URA scheme using exhaustive computer simulations. Finally, we draw out some concluding remarks in Section V.

We also mention the common notations used in this paper. 
 Lower- and upper-case bold fonts, $\mathbf{x}$ and $\mathbf{X}$, are used to denote vectors and matrices, respectively. Upper-case calligraphic font, $\mathcal{X}$ and $\bm{\mathcal{X}}$, is used to denote single and multivariate random variables, respectively, as well as for sets notation (depending on the context). The $(m,n)$th entry of $\mathbf{X}$ is denoted as  $\mathbf{X}_{mn}$, and the $n$th element of $\mathbf{x}$ is denoted as $x_n$.
    The identity matrix is denoted as $\mathbf{I}$. The operator $\textrm{vec}(\mathbf{X})$ stacks the columns of a matrix $\mathbf{X}$ one below the other. The shorthand notation $\bm{\mathcal{X}} \sim\mathcal{CN}(\mathbf{x};\mathbf{m},\mathbf{R})$ means that the random vector $\bm{\mathcal{X}}$ follows a complex circular Gaussian distribution with mean $\mathbf{m}$ and auto-covariance matrix $\mathbf{R}$. Likewise, $\mathcal{X}\sim\mathcal{N}(x;m,\mu)$ means that the random variable $\mathcal{X}$ follows a Gaussian distribution with mean $m$ and variance $\mu$. Moreover, $\{.\}^\textsf{T}$ and $\{.\}^\textsf{H}$ stand for the transpose and Hermitian (transpose conjugate) operators, respectively. In addition, $|.|$ and $\|.\|$ stand for the modulus and Euclidean norm, respectively. Given any complex number,  $\Re\{.\}$, $\Im\{.\}$, and $\{.\}^*$ return its real part, imaginary part, and complex conjugate, respectively. The Kronecker  function and product are denoted as $\delta_{m,n}$ and $\otimes$, respectively.  We also denote the probability distribution function (pdf) of single and multivariate random variables (RVs) by $p_{\mathcal{X}}(x)$ and $p_{\bm{\mathcal{X}}}(\mathbf{x})$, respectively. The statistical expectation is denoted as $\mathbb{E}\{.\}$, $j$ is the imaginary unit (i.e., $j^{2}=-1$), and the notation $\triangleq$ is used for definitions.

\section{System Model and Assumptions}\label{section_2}

Consider a single-cell network consisting of $K$ single-antenna devices which are being served by a base station located at the center of a cell of radius $R$. Devices are assumed to be uniformly scattered inside the cell, and we denote by $r_k$ (measured in meters) the distance from the $k$th device to the base station. This paper assumes sporadic device activity thereby resulting in a small number, $K_a\ll K$, of devices being active over each coherence block. The devices communicate to the base station through the uplink uncoordinated scheme, in which every active device wishes to communicate $B$ bits of information over the channel in a single communication round. The codewords transmitted by active devices are drawn uniformly from a common Gaussian codebook $\mathcal{C}=\big\{\widetilde{\mathbf{c}}_1, \widetilde{\mathbf{c}}_2, \cdots,\widetilde{\mathbf{c}}_{2^B} \big\} \subset \mathbb{C}^{n}$. More precisely, $\widetilde{\mathbf{c}}_b\sim\mathcal{CN}(\mathbf{0},P_{t}\mathbf{I})$ where $n$ is the blocklength and $P_t$ is the transmit power. We model the device activity and codeword selection by a set of $2^B K$ Bernoulli random variables $\delta_{b,k}$ for $k=1,...,K$ and $b=1,...,2^B$
\begin{eqnarray}
\delta_{b,k}&=&\left\{\begin{array}{ll}
{1} & \text {if user $k$}  \text { is active and~} \text{transmits codeword}~ \widetilde{\mathbf{c}}_b,\\
{0} & {\text {otherwise. }  \text {  }}
\end{array}\right.\nonumber
\end{eqnarray}

We consider a Gaussian multiple access channel (MAC) with a block fading model and a large-scale antenna array consisting of $M_r$ receive antenna elements at the BS.  Assuming the channels remain almost unchanged over the entire transmission period, the uplink received signal at the $m$th antenna element can be expressed as follows:
\begin{equation}\label{wireless:two_sums_1}
\widetilde{\mathbf{y}}^{(m)}~=~ \sum_{k=1}^{K}\sum_{b=1}^{2^B}\sqrt{g_{k}}\widetilde{h}_{k,m}\delta_{b,k}\widetilde{\mathbf{c}}_{b}~+~{\widetilde{\mathbf{w}}^{(m)}},~~  m = 1,\ldots,M_r.
\end{equation}
 The random noise vector, $\widetilde{\mathbf{w}}^{(m)}$, is modeled by a complex circular Gaussian random vector with independent and identically distributed (i.i.d.) components, i.e., $\widetilde{\mathbf{w}}^{(m)}\sim\mathcal{CN}(\mathbf{0},\mathbf{\sigma}_w^2\mathbf{I})$. In addition, $\widetilde{h}_{k,m}$ stands for the small-scale
fading coefficient between the $k$th user and the $m$th antenna. We assume Rayleigh block fading, i.e., the small-scale fading channel coefficients, $\widetilde{h}_{k,m}\sim\mathcal{CN}(0,1)$, remain constant over the entire observation window which is smaller than the coherence time. Besides, $g_k$ is the large-scale fading coefficient of user $k$ given by (in dB scale):
\begin{eqnarray}\label{wireless:large-scale_fading} 
g_{k}\,[\mathrm{d} \mathrm{B}]~=~-\alpha-10\beta \log _{10}\left(r_{k}\right),
\end{eqnarray}
where $\alpha$ is the fading coefficient measured at distance $d=5 ~$ meter and $\beta$ is the pathloss exponent. For convenience, we also define the effective channel coefficient by lumping the large- and small-scale fading coefficients  together in one quantity, denoted as $h_{k,m}\triangleq\sqrt{g_k}\widetilde{h}_{k,m}$, thereby yielding the following equivalent model:
\begin{eqnarray}\label{wireless:system_model_equiv}
\widetilde{\mathbf{y}}^{(m)}~=~ \sum_{k=1}^{K}\sum_{b=1}^{2^B}h_{k,m}\delta_{b,k}\widetilde{\mathbf{c}}_{b}~+~{\widetilde{\mathbf{w}}^{(m)}},~~  m = 1,\ldots,M_r.
\end{eqnarray} 

To define the random access code for this channel, let $W_k \in ~[2^B]~\triangleq \{1,2,\ldots,2^B\}$ denote the message of user $k$, such that for some encoding function $f:[2^B] \rightarrow \mathbb{C}^{n}$, we have $f(W_k) = \widetilde{\mathbf{c}}_{b_k}$. By recalling that $K_a$ stands for the number of active users, the decoding\footnote{The notation ${{[2^B]}\choose {K_{a}}}$ stands for choosing $K_a$ different elements from the set $[2^B]$.} map $g: \mathbb{C}^{n\times M_r} \rightarrow{{[2^B]}\choose {K_{a}}}$ outputs a list of $K_a$ decoded messages with the probability of error being defined as: 
\begin{eqnarray}\label{wireless:error_prob}
 P_e ~=~\frac{1}{K_a}\sum_{k=1}^{K_a}\Pr(E_k),
\end{eqnarray}  
and $E_{k}\triangleq\big\{W_{k} \notin g({\widetilde{\mathbf{y}}^{(1)}, \widetilde{\mathbf{y}}^{(2)},\ldots ,\widetilde{\mathbf{y}}^{(M_r)}})\big\}$. Notice here that  $P_e$ depends solely on the number of active users, $K_a$, instead of the total number of users $K$.
With this formulation in mind,  we rewrite (\ref{wireless:system_model_equiv}) in a more succinct matrix-vector form as follows:
\begin{eqnarray}\label{wireless:Delta}
\widetilde{\mathbf{y}}^{(m)}~=~\widetilde{\mathbf{C}}\mathbf{\widetilde{\Delta}}\mathbf{h}^{(m)}~+~\widetilde{\mathbf{w}}^{(m)},~~  m = 1,\ldots,M_r,
\end{eqnarray} 
in which $\widetilde{\mathbf{C}}=\big[\widetilde{\mathbf{c}}_1, \widetilde{\mathbf{c}}_2, \ldots,\widetilde{\mathbf{c}}_{2^B} \big] \in\mathbb{C}^{n\times2^B}$ is the codebook matrix, which is common to all the users and   $\mathbf{h}^{(m)}=[h_{1,m},h_{2,m},\ldots,h_{K,m}]^\textsf{T}\in\mathbb{C}^{K}$ is the multi-user channel vector at
the $m$th antenna which incorporates the small- and large-scale fading coefficients. The  matrix  $\widetilde{\mathbf{\Delta}}\in\{0,1\}^{2^B\times K}$ contains only $K_a$ non-zero columns each of which having a single non-zero entry. Observe here that both $\widetilde{\mathbf{\Delta}}$ and $\mathbf{h}^{(m)}$ are  unknown to the receiver. 
Hence, by defining $\widetilde{\mathbf{x}}^{(m)} \triangleq \widetilde{\mathbf{\Delta}}{\mathbf{h}}^{(m)}$, it follows that:
\begin{eqnarray}\label{wireless:system_model_equiv_mtx}
\widetilde{\mathbf{y}}^{(m)}~=~\widetilde{\mathbf{C}}\,\widetilde{\mathbf{x}}^{(m)}~+~\widetilde{\mathbf{w}}^{(m)},~~  m = 1,\ldots,M_r.
\end{eqnarray}

Note here that each active user contributes a single non-zero coefficient in $\widetilde{\mathbf{x}}^{(m)}$ thereby resulting in  $K_a-$sparse $2^B-$dimensional   vector. Since $K_a$ is much smaller than the total number of codewords $2^B$, $\widetilde{\mathbf{x}}^{(m)}$ has a very small  sparsity ratio $\lambda\triangleq\frac{K_a}{2^B}$. Observe also that the formulation in (\ref{wireless:system_model_equiv_mtx}) belongs to the MMV class in compressed sensing terminology, which can be equivalently rewritten in a more succinct matrix/matrix form as follows:
\begin{eqnarray}\label{wireless:system_model_mtx_MMV}
\widetilde{\mathbf{Y}} ~=~ \widetilde{\mathbf{C}}\widetilde{\mathbf{X}}~+~\widetilde{\mathbf{W}},
\end{eqnarray}
in which $\widetilde{\mathbf{Y}} = \big[\widetilde{\mathbf{y}}^{(1)}, \widetilde{\mathbf{y}}^{(2)},...,\widetilde{\mathbf{y}}^{(M_r)}\big]$ is the entire measurement matrix and $\widetilde{\mathbf{X}} = \big[\widetilde{\mathbf{x}}^{(1)}, \widetilde{\mathbf{x}}^{(2)},...,\widetilde{\mathbf{x}}^{(M_r)}\big]$. With this formulation, the unknown matrix $\widetilde{\mathbf{X}}$ is row-sparse and we aim to exploit this structure by casting our task into the problem of   estimating a group-sparse vector from a set of linear measurements. The theory of sparse reconstruction from noisy observations has been a hot research topic in statistics and we refer the theoretically inclined reader to chap.  7-9 in \cite{wainwright2019high}, for elaborate discussions on the theoretical guarantees for sparse reconstruction. 

\section{Proposed  Unsourced Random Access Scheme Based on Compressive Sensing  }\label{section_3}
\subsection{Slotted Transmission Model and Code Selection}

By revisiting (\ref{wireless:two_sums_1}), we see that the number of codewords  grows exponentially with the blocklength $n$. Indeed, for a fixed rate $R = \frac{B}{n}$, we have $2^B = 2^{nR}$ --- which becomes extremely large even at moderate values of $n$ --- thereby making any attempt to directly use standard sparse recovery algorithms computationally prohibitive. Practical approaches have been introduced to alleviate this computational burden, including the slotted transmission framework also adopted in this paper. Indeed, similar to \cite{5G_3}, each active user  partitions its $B-$bit message into $L$ equal-size  information bit sequences (or chunks). As opposed to \cite{5G_3}, however, our approach does not require concatenated coding to couple the sequences across different slots (i.e., no outer binary encoder). Therefore, we simply share the bits uniformly between the $L$ slots and there is no need to optimize  the  sizes of the $L$ sequences. In this way, there is a total number of $J = \frac{B}{L}$ bits in each sequence (i.e., associated to each slot). 

Let the matrix $\widetilde{\mathbf{A}}\in\mathbb{C}^{\frac{n}{L}\times 2^J}$ denote the common codebook for all the users (over all slots). That is, the columns of $\widetilde{\mathbf{A}} = [\widetilde{\mathbf{a}}_1,\widetilde{\mathbf{a}}_2,\ldots,\widetilde{\mathbf{a}}_{2^J}]$ form a set of codewords that each $\{k^{th}\}_{k=1}^{K_a}$ active user chooses from in order to encode its $\{l^{th}\}_{l=1}^L$ sequence before transmitting it over the $\{l^{th}\}_{l=1}^L$ slot. Notice here that, in such a slotted transmission framework, the size of the codebook  is ${2^J}$. This is much smaller than the original codebook size, ${2^B}$, which was actually used to prove the random coding achievability bound in \cite{5G_1}. Slotting is, however, a necessary step towards alleviating the computational burden as mentioned previously. Yet, it is still essential for our proposed scheme to choose a sufficiently large value for $J$ such that the expected number of collisions remains small as compared to the number of  users. More specifically, from the union bound estimate, it necessary that  $\frac{2L{{K_a}\choose {2}}}{K_a2^J}$ does not dominate the probability of incorrect decoding. The simulation results in Section \ref{section_5} suggest that $J=17$ is enough to keep the contribution of collision-induced errors to the overall error probability negligible.
\subsection{Encoding}
\begin{figure}[h!]
    \centering
    \includegraphics[width=1\linewidth]{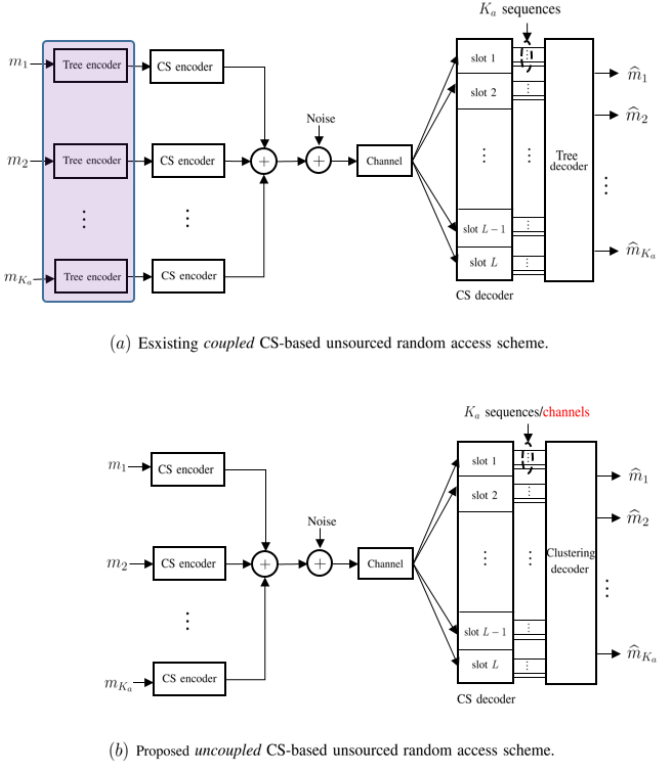} 
    \vskip 0.5cm
    \caption{High-level description of the (existing) coded/coupled and the (proposed) uncoded/uncoupled CS-based unsourced random access schemes. The main differences lye in $i)$ removing the outer tree encoder which is highlighted in purple colour in the top figure and $ii)$ replacing the computationally intensive outer tree decoder by a simple clustering-type decoder.}  
    \label{fig:arcitecture}
\end{figure}

After partitioning each packet/message into $L$ $J-$bit information sequences, the latter are encoded separately using the codebook, $\widetilde{\mathbf{A}} \in \mathbb{C}^{\frac{n}{L}\times 2^J}$, which will serve as the sensing matrix for sparse recovery.  Conceptually, we operate on a per-slot basis by associating to every possible $J-$bit information sequence a different column in the codebook  matrix $\widetilde{\mathbf{A}}$. Thus, we can view this matrix as a set of potentially transmitted messages over the duration of a slot. The  multiuser CS encoder can be visualized as an abstract multiplication of  $\widetilde{\mathbf{A}}$ by an index vector $\mathbf{v}$. The positions of non-zero coefficients in $\mathbf{v}$ are nothing but the decimal representations of the information bit sequences/chunks being transmitted by the active users over a given slot. Thus, the slotted transmission of the $B$-bit packets of all the active users gives rise to $L$ small-size compressed sensing instances (one per each slot). Now, after encoding its $J-$bit sequence, user $k$ modulates the corresponding codeword and transmits it over the channel where it is being multiplied by a complex coefficient $h_{k,m}$ before reaching the $m$th antenna. Hence, the overall baseband model over each slot reduces to the MAC model discussed in Section \ref{section_2}. Hence, by recalling (\ref{wireless:system_model_mtx_MMV}), the received signal over the $l$th slot is given by:
\begin{eqnarray}\label{wireless:eqn_per_slot_1}
\widetilde{\mathbf{Y}}_l ~=~ {\widetilde{\mathbf{A}}}\widetilde{\mathbf{X}}_l~+~\widetilde{\mathbf{W}}_l, ~~  l = 1,\ldots,L.
\end{eqnarray}
Vectorizing (\ref{wireless:eqn_per_slot_1}) yields:
\begin{eqnarray}\label{sireless:system_model_vec} 
\textrm{vec}\big(\widetilde{\mathbf{Y}}_l^{\textsf{T}}\big) ~=~ (\widetilde{\mathbf{A}}^{\textsf{T}}\otimes \mathbf{I})\textrm{vec}\big(\widetilde{\mathbf{X}}_l^{\textsf{T}}\big)~+~\textrm{vec}\big(\widetilde{\mathbf{W}}_l^{\textsf{T}}\big),
\end{eqnarray} 
in which $\otimes$ denotes the Kronecker product of two matrices. Then, by defining $\widetilde{\bm{\mathbb{A}}}\,\triangleq\,\widetilde{\mathbf{A}}^{\textsf{T}}\otimes \mathbf{I}\in\mathbb{C}^{\frac{n}{L}M_r\times 2^JM_r}$, $\widetilde{{\mathbf{y}}}_l\,\triangleq\,\textrm{vec}\big(\widetilde{\mathbf{Y}}_l^{\textsf{T}}\big)\in\mathbb{C}^{\frac{n}{L}M_r}$, $\widetilde{\mathbf{x}}_l\,\triangleq\,\textrm{vec}\big(\widetilde{\mathbf{X}}_l^{\textsf{T}}\big)\in\mathbb{C}^{2^JM_r}$, and $\widetilde{\mathbf{w}}_l\,\triangleq\,\textrm{vec}\big(\widetilde{\mathbf{W}}_l^{\textsf{T}}\big)$,
we recover the problem of estimating a sparse vector, $\bar{\mathbf{h}}_l$, from its noisy linear observations:
\begin{eqnarray}\label{wireless:eqn_per_slot}
\widetilde{\mathbf{y}}_l ~=~ \widetilde{\mathbb{A}}\, \widetilde{\mathbf{x}}_l~+~\widetilde{\mathbf{w}}_l, ~~  l = 1,\ldots,L. 
\end{eqnarray} 
Fig. \ref{fig:arcitecture} schematically depicts the proposed URA scheme and contrasts it to the existing coded/coupled compressed sensing-based  scheme. As seen there, the proposed scheme eliminates the need for concatenated coding, i.e., the outer tree encoder.
Indeed, instead of coupling the slot-wise information sequences through additional parity-check bits to be able to stitch them at the receiver, the proposed scheme leverages the inherent coupling provided by nature in the form of channel correlations across slots. In other words, if one is able to find the assignment matrix that clusters the slot-wise reconstructed channels for each user together, then the decoded sequences can also be clustered (i.e., stitched) in the same way.
To better reconstruct the channels, this paper postulates a Bernoulli-Laplacian distribution as a heavy-tailed prior on $h_{k,m}$. The rationale for this choice will be discussed in some depth in Section \ref{Part_C}. Since the Laplacian distribution is defined for real-valued random variables only, we transform the complex-valued model in (\ref{wireless:eqn_per_slot}) into its equivalent real-valued model as follows:
\begin{eqnarray}\label{transfromed_model}
     \underbrace{\begin{bmatrix}
       \Re\{\widetilde{\mathbf{y}}_l\}\, \\[0.4em]      
      \Im\{\widetilde{\mathbf{y}}_l\}
     \end{bmatrix}}_{\substack{\\\mathlarger{\triangleq}\\\mathlarger{\mathbf{y}_l}}}
     &\!\!=\!\!& 
     \underbrace{\begin{bmatrix}
       \Re\{\bar{\mathbb{A}}\} & -\Im\{\bar{\mathbb{A}}\}          \\[0.4em]
       \Im\{\bar{\mathbb{A}}\}          & \Re\{\bar{\mathbb{A}}\}
     \end{bmatrix}}_{\substack{\\\mathlarger{\triangleq}\\\mathlarger{\bar{\mathbf{A}}}}} 
     \underbrace{
     \begin{bmatrix}
       \Re\{\widetilde{\mathbf{x}}_l\} \\[0.4em]      
      \Im\{\widetilde{\mathbf{x}}_l\}
     \end{bmatrix}
     }_{\substack{\\\mathlarger{\triangleq}\\\mathlarger{\bar{\mathbf{x}}_l}}}
     +\underbrace{
     \begin{bmatrix}
       \Re\{\widetilde{\mathbf{w}}_l\} \\[0.4em]      
       \Im\{\widetilde{\mathbf{w}}_l\}
     \end{bmatrix}
     }_{\substack{\\\mathlarger{\triangleq}\\\mathlarger{\mathbf{w}_l}}}.\nonumber\\
\end{eqnarray} 

 Finally, by defining $M \triangleq \frac{2nM_r}{L}$, and $N \triangleq 2M_r2^J$, the goal is to reconstruct the unknown sparse vector, $\bar{\mathbf{x}}_l\in\mathbb{R}^{N}$, given by:
\begin{align}\label{slotwisevector}
 \bar{\mathbf{x}}_l = \Big[\Re\{\widetilde{\mathbf{x}}_{1,l}\}^{\textsf{T}},\ldots,\Re\{\widetilde{\mathbf{x}}_{2^J,l}\}^{\textsf{T}},\Im\{\widetilde{\mathbf{x}}_{1,l}\}^{\textsf{T}},\ldots,\Im\{\widetilde{\mathbf{x}}_{2^J,l}\}^{\textsf{T}}\Big]^{\textsf{T}}
\end{align}
based on the knowledge of $\mathbf{y}_l\in\mathbb{R}^{M}$ and 
$\bar{\mathbf{A}}\in\mathbb{R}^{M \times N}$. We emphasize here the fact that $\bar{\mathbf{x}}_l$  
has a block-sparsity structure with dependent blocks since whenever $\Re\{\widetilde{\mathbf{x}}_{j,l}\}\,=\,\mathbf{0}$  then $\Im\{\widetilde{\mathbf{x}}_{j,l}\}\,=\,\mathbf{0}$. 

For ease of exposition, we slightly rewrite (\ref{transfromed_model}) to end up with a convenient model in which we are interested in reconstructing the following group sparse vector: 

\begin{eqnarray}
  \mathbf{x}_l=\Big [\underbrace{\Re\{\widetilde{\mathbf{x}}_{1,l}\}^{\textsf{T}},\Im\{\widetilde{\mathbf{x}}_{1,l}\}^{\textsf{T}}}_{  \mathbf{x}_{1,l}},\ldots,\underbrace{\Re\{\widetilde{\mathbf{x}}_{2^J,l}\}^{\textsf{T}},\Im\{\widetilde{\mathbf{x}}_{2^J,l}\}^{\textsf{T}}}_{  \mathbf{x}_{2^J,l}}\Big]^{\textsf{T}},
\end{eqnarray}

which has independent sparsity among its constituent blocks $\{\mathbf{x}_{j,l}\}_{j=1}^{2^J}$. To achieve this, observe that $\mathbf{x}_l$ and $\bar{\mathbf{x}}_l$ are related as follows:
\begin{eqnarray}\label{permuted_x}
 \mathbf{x}_l~=~\overline{\bm{\Pi}}\,\bar{\mathbf{x}}_l,  
\end{eqnarray}
for some know permutation matrix, $\overline{\bm{\Pi}}$, which satisfies $\overline{\bm{\Pi}}^{\textsf{T}}\,\overline{\bm{\Pi}}=\mathbf{I}$. By plugging (\ref{permuted_x}) in (\ref{transfromed_model}), we obtain the following equivalent CS problem: 
\begin{eqnarray}\label{CS_Slot_l}
 \mathbf{y}_l~=~\mathbf{A}\,\mathbf{x}_l~+~\mathbf{w}_l~~~~\textrm{with}~~~~\textbf{A}~\triangleq~\bar{\textbf{A}}\overline{\bm{\Pi}}^{\textsf{T}}.
\end{eqnarray}

\subsection{\changef{CS} Recovery and Clustering-Based Stitching:}\label{Part_B}
The ultimate goal at the receiver is to identify the set of $B$-bit messages that were transmitted by all the active users. Since the messages were partitioned into $L$ different chunks, we obtain an  instance of unsourced MAC in each slot. The inner CS-based decoder must now decode, in each slot, the  $J-$bit sequences of all the $K_a$ active users. The outer clustering-based decoder 
will put  together the slot-wise decoded sequences of each user, so as to recover all the original transmitted $B$-bit messages (cf. Fig. \ref{fig:arcitecture} for more details).

	For each $l$th slot, the task is then to first reconstruct ${\mathbf{x}}_l$ from ${\mathbf{y}}_l ~=~ \mathbf{A}{\mathbf{x}}_l~+~{\mathbf{w}}_l$ given ${\mathbf{y}}_l$ and ${\mathbf{A}}$. To solve the joint activity detection and channel estimation problem, we resort to the HyGAMP CS algorithm  \cite{Hybrid_Gamp} and we will also rely on the EM-concept \cite{dempster1977maximum} to learn the unknown hyperparameters of the model. In particular, we embed the EM algorithm inside HyGAMP  to learn  the variances of the additive noise and the postulated prior, which are both required to execute HyGAMP itself. HyGAMP makes use of  large-system Gaussian and quadratic  approximations for the messages of loopy belief propagation on the factor graph. As opposed to GAMP \cite{GAMP}, HyGAMP is able to accommodate the group sparsity structure in $\mathbf{x}_l$ by using a dedicated latent Bernoulli random variable, $\varepsilon_j$, for each $\{j^{th}\}_{j=1}^{2^J}$ group, $\mathbf{x}_{j,l}$,  in $\mathbf{x}_l$. We will soon see how HyGAMP finds the MMSE and MAP estimates,  $\{\widehat{\mathbf{x}}_{j,l}\}_{j=1}^{2^J}$ and $\{\widehat{\epsilon}_j\}_{j=1}^{2^J}$ of $\{\mathbf{x}_{j,l}\}_{j=1}^{2^J}$ and $\{{\varepsilon}_j\}_{j=1}^{2^J}$. The latter will be in turn used to decode the transmitted sequences in each slot (up to some unknown permutations) while by clustering the MMSE estimates of the active users' channels it is possible to recover those unknown permutations and correctly stitch the decoded sequences.
For this reason, we denote the $K_a$ reconstructed channels over each $l$th slot (i.e., the nonzero blocks in the entire reconstructed vector $\widehat{\mathbf{x}}_l\,=\,\big[\widehat{\mathbf{x}}_{1,l},\widehat{\mathbf{x}}_{2,l},..., \widehat{\mathbf{x}}_{2^j,l}\big]^{\textsf{T}}$ as  $\{\widehat{\mathbf{h}}_{k,l}\}_{k=1}^{K_a}$. By denoting, the residual estimation noise as $\mathbf{\widehat{\mathbf{w}}}_{k,l}$, it follows that:    
\begin{eqnarray}\label{reconstructed_channels}
\widehat{\mathbf{h}}_{k,l}~=~\bar{\mathbf{h}}_k~+~\mathbf{\widehat{\mathbf{w}}}_{k,l},~~ k = 1,\ldots, K_a,~l=1,...,L,
\end{eqnarray}
in which $\bar{\mathbf{h}}_k\,\triangleq\,\big[\Re\{\mathbf{h}_{k}\},\Im\{\mathbf{h}_{k}\}\big]^{\textsf{T}}$ with $\mathbf{h}_{k}\,\triangleq\,\big[h_{k,1}, h_{k,2},...,h_{k,M_r}\big]^{\textsf{T}}$ is the true complex channel vector for user $k$. 
The outer clustering-based decoder takes the $LK_a$   reconstructed channels in (\ref{reconstructed_channels}) --- which are slot-wise permuted --- and returns one cluster per active user, that contains its $L$ noisy  channel estimates.

 To cluster the reconstructed channels into $K_a$ different groups, we resort to the  Gaussian mixture expectation-maximization  procedure which consists of fitting a Gaussian mixture distribution to the data points in (\ref{reconstructed_channels}) under the assumption of Gaussian residual noise. We also assume the reconstruction noise to be Gaussian which is a common practice in the approximate message passing framework including HyGAMP.    However, as seen from (\ref{transfromed_model}) the matrix $\bar{\mathbf{A}}$ is not i.i.d Gaussian as would be required to rigorously prove the above claim which is a widely believed conjecture based on the concept of universality from statistical physics \cite{tulino2013support},\cite{abbara2020universality}.     Moreover, we will devise an appropriate constrained clustering procedure that enforces the following two constraints that are very specific to our problem: $i)$  each cluster must have exactly $L$ data points, and $ii)$ channels reconstructed over the same slot must not be assigned to the same cluster.  

\subsection{Hybrid Approximate Message Passing }\label{Part_C}
In this section, we describe the HyGAMP CS algorithm \cite{Hybrid_Gamp} by which we estimate the channels and decode the data in each slot. As a matter of fact, decoding the transmitted messages in slot $l$ comes as a byproduct of reconstructing the entire group-sparse vector ${\mathbf{x}}_l$. This is because there is a one-to-one mapping between the positions of the non-zero blocks in  $\mathbf{x}_l$ and the transmitted codewords that are drawn from the common codebook $\widetilde{\mathbf{A}}$. As mentioned earlier, HyGAMP finds asymptotic MMSE estimates for the entries of the group-sparse vector, ${\mathbf{x}}_l$, in each slot $l$. To capture the underlying group sparsity structure, HyGAMP uses the following set of Bernoulli latent random variables:
\begin{eqnarray}\label{prior_epsilon}
\varepsilon_{j}=\left\{\begin{array}{ll}
{1} & {\text { if group } j \text { is active}}, \\ 
{0} & {\text { if group } j \text { is inactive}},
\end{array}\right.
\end{eqnarray}
which are i.i.d with the common prior  $\lambda ~\triangleq~ \Pr(\varepsilon_j = 1) ~=~ \frac{K_a}{2^J}$.   The marginal posterior probabilities, $\Pr(\varepsilon_j = 1\mid\mathbf{y}_l)$, for $j=1,2,\ldots,2^J$ are given by:
\begin{eqnarray}\label{posterior} 
\Pr(\varepsilon_j = 1\mid\mathbf{y}_l) ~=~ \frac{\lambda}{\lambda+ (1-\lambda)\exp\left(\displaystyle-\sum_{q=1}^{2M_r}\textrm{LLR}^{(l)}_{q\rightarrow j }\right)},
\end{eqnarray}
where $\textrm{LLR}^{(l)}_{q\rightarrow j }$ is updated in line 19 of Algorithm 1 (cf. next page) while trying to reconstruct the unknown  channels\footnote{ For ease of notation we drop the slot index $l$ in Algorithm 1.}. The posterior probabilities, $\{\Pr(\varepsilon_j = 1\mid\mathbf{y}_l)\}_{j=1}^{2^J}$, are used by the receiver to infer which of the codewords were transmitted by the active users over slot $l$. This is done by simply returning the columns in $\widetilde{\mathbf{A}}$ that correspond to the $K_a$ largest values among the posterior probabilities in (\ref{posterior}).

\begin{figure}[h!]
\hskip -0.4cm  
    \includegraphics[scale=0.56]{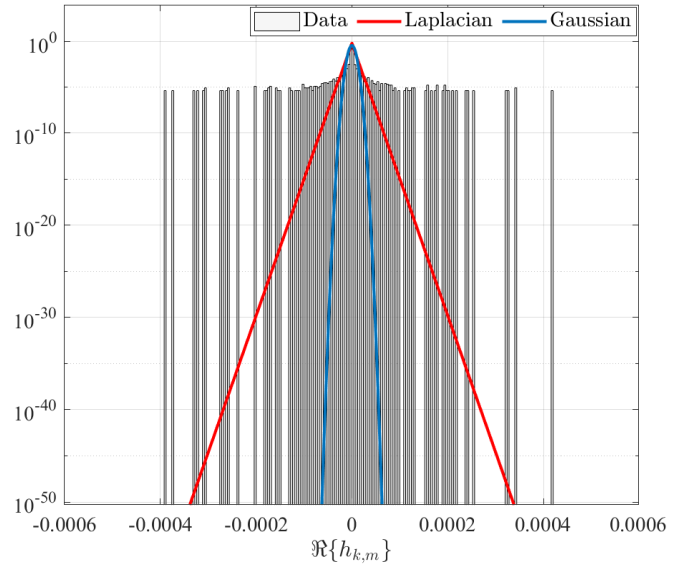}
    \caption{Histogram of the channel coefficients (real part) together with the Laplacian and Gaussian PDFs fitted to it.}
    \label{fig:histogram_simulation} 
\end{figure}

Note here that for the sake of simplicity, we assume the number of active users, $K_a$, to be known to the receiver as is the case in all existing works on unsourced random access. Yet, we emphasize the fact that it is straightforward to generalize our approach to also detect the number of active users by learning the hyperparameter $ \lambda = \frac{K_a}{2^J}$ using the EM procedure as done in \cite{Schniter_EM_GM}. Motivated by our recent results in \cite{bellili2019generalized}, we also postulate a Bernoulli-Laplacian prior to model the channel coefficients. The main rationale behind choosing this prior is the need for using a heavy-tailed distribution to capture the effect of the large-scale fading coefficients, $\sqrt{g_k}$, which vary drastically  depending on the relative users' locations with respect to the BS.  
Indeed, unlike most AMP-based works on massive activity detection (e.g., \cite{liu2018massive}) which assume perfect knowledge of the large-scale fading coefficients, in our paper the latter are absorbed in the the overall channel coefficients and estimated with them using HyGAMP. Therefore, we had to opt for a heavy-tailed prior to capture the rare events of getting an active user close to the base station and whose channel will be very large compared to most of the other faraway active users. In this respect, the Bernoulli-Laplacian prior was found to offer a good trade-off between denoising difficulty and model-mismatch. The Bernoulli-Laplacian is also computationally more attractive than other heavy-tailed priors since it requires updating only one parameter using the nested EM algorithm (inside HyGAMP) as will be explained later on. As an empirical evidence, in Fig. \ref{fig:histogram_simulation},  we plot  the Laplacian distribution:
  \begin{eqnarray}\label{laplacian_prior}
 \mathcal{L}(x;\sigma_x) = \frac{1}{2\sigma_x} e^{-\frac{|x|}{\sigma_x}},
\end{eqnarray}
 with $\sigma_x = \sqrt{\sigma^2/2}$ wherein $\sigma^2$ is the empirical variance  of the data that is extracted from active users' channels only, i.e., $\Re\{h_{k,m}\} = \sqrt{g_{k}}\Re\{\widetilde{h}_{k,m}\}$. Fig. \ref{fig:histogram_simulation} also depicts the Gaussian pdf  after fitting it  to the same data  set. There, it is seen that the Laplacian prior exhibits a better fit to the data and we noticed that owing to its heavier tail it enables HyGAMP to better capture both cell-center and cell-edge users.
 
\begin{algorithm*}[!ht]
   \begin{algorithmic}[1]\vskip 0.1cm
  
    \Require
      $\mathbf{A}\in \mathbb{R}^{M\times N}$; $\mathbf{y}\in\mathbb{R}^{M}$; $\sigma_x$, $\lambda$, $\sigma_w^2$, precision tolerance ($\xi$), maximum number of iterations ($T_\textsc{max}$)\vskip 0.2cm
    \Ensure
      MMSE estimates, $\{\widehat{x}_{q,j}\}_{q=1}^{2M_r}$, of $\{x_{q,j}\}_{q=1}^{2M_r}$~ $\forall j$, and MAP estimates, $\{\widehat{\varepsilon}_j\}_{j=1}^{2^J}$, of $\{\varepsilon_j\}_{j=1}^{2^J} $  \\ \vskip 0.3cm  
     \textbf{Initialization}
     \State $t\gets 0$
           \State $\forall q,j:~\mu_{q,j}^r(t-1)\, =\, 1$
       \State $\forall q,j:~\widehat{r}_{q,j}(t-1) \,=\, 0$
       \State $\forall q,j:~\mathrm{LLR}_{q\leftarrow j}(t-1) \,=\, \log (\lambda /(1-\lambda))$\label{algo:LLR-init}
      \State  $\forall q,j:~{\widehat{\rho}}_{q,j}(t) \,=\,  1 /\left[1+\exp\big(- \mathrm{LLR}_{q\leftarrow j}(t-1)\big)\right]$\label{algo:sparsity-init}
    \Repeat 
    \State $\forall q,j:~\widehat{x}_{q,j}(t) ~=~ \mathbb{E}_{\mathcal{X}_{q,j}|\bm{\mathcal{Y}}}\Big\{x_{q,j}\big|\mathbf{y}\,;\,\widehat{r}_{q,j}(t-1),\mu^r_{q,j}(t-1),\widehat{\rho}_{q,j}(t),\sigma_x\Big\}$\vskip 0.1cm\label{algo:mean_xn}
     \State  $\forall q,j:~\mu^x_{q,j}(t)~ =~ \textsf{var}_{\mathcal{X}_{q,j}|\bm{\mathcal{Y}}}\Big\{x_{q,j}\big|\mathbf{y}\,;\,\widehat{r}_{q,j}(t-1),\mu^r_{q,j}(t-1),\widehat{\rho}_{q,j}(t),\sigma_x\Big\}$\vskip 0.1cm\label{algo:var_xn}
     \State  $\forall i$:~ $\widehat{z}_i(t) = \sum_{q,j}\mathbf{A}_{iq}^{(j)}\widehat{x}_{q,j}(t)$\vskip 0.1cm
\State $\forall i:~\mu_i^p(t) ~=~ \sum_{q,j}|\mathbf{A}_{iq}^{(j)}|^2\mu_{q,j}^x(t)$\vskip 0.1cm\label{algo:var_gauss_out}
\State $\forall i:~\widehat{p}_i(t)\, =\, \widehat{z}_i(t)\,-\,\mu_i^p(t)\widehat{s}_i(t-1)$\vskip 0.1cm
\State $\forall i:~\widehat{z}_i^0(t) ~=~ \mathbb{E}_{\mathcal{Z}_i|\bm{\mathcal{Y}}}\Big\{z_i\big|\mathbf{y}\,;\,\widehat{p}_i(t),\mu^p_i(t),\sigma_w^{2}\Big\}$\vskip 0.1cm\label{algo:mean_zm}
\State $\forall i:~\mu_i^z(t) ~=~ \textsf{var}_{\mathcal{Z}_i\big|\bm{\mathcal{Y}}}\Big\{z_i|\mathbf{y}\,;\,\widehat{p}_i(t),\mu^p_i(t),\sigma_w^{2}\Big\}$\vskip 0.1cm\label{algo:var_zm}
\State $\forall i:~\widehat{s}_i(t) ~=~ \frac{1}{\mu_i^p(t)}\big[\widehat{z}_i^0(t)-\widehat{p}_i(t)\big]$\vskip 0.1cm\label{algo:var_gauss_in}
\State $\forall i:~\mu_i^s(t)\, =\, \frac{1}{\mu_i^p(t)}\left[1-\frac{\mu_i^z(t)}{\mu_i^p(t)}\right]$\vskip 0.1cm
\State $\forall q,j:~\mu_{q,j}^r(t) ~=~ \left(\sum_{i}|\mathbf{A}_{iq}^{(j)}|^2\mu_i^s(t)\right)^{-1}$\vskip 0.1cm\label{algo:var_gauss_out1}
     \State $\forall q,j:~\widehat{r}_{q,j}(t)~ =~ \widehat{x}_{q,j}(t)~+~ \mu^r_{q,j}(t)\sum_{i}\mathbf{A}^{(j)}_{iq}\widehat{s}_i(t)$\vskip 0.1cm
    \State  $\forall q,j:\text { Compute } \textrm{LLR}_{q \rightarrow  j}(t) ~\text{using (\ref{LLR_update})  }$\vskip 0.1cm\label{algo:LLR_reverse}
    \State $\forall q,j~:\textrm{LLR}_{q\leftarrow j}(t) \,=\, \log (\lambda /(1-\lambda))+\sum_{q' \neq q} \textrm{LLR}_{q' \rightarrow j}(t)$\vskip 0.1cm\label{algo:LLR}
    \State $\forall q,j~:{\widehat{\rho}}_{q,j}(t+1) \,=\, 1/\left[1+ \exp \big(-\textrm{LLR}_{q \leftarrow j}(t)\big)\right]$\label{algo:sparsity_param}
       \State  $t\gets t+1$\vskip 0.3cm 
    \Until{$\big|\!\big|\widehat{\mathbf{x}}(t+1)\,-\,\widehat{\mathbf{x}}(t)\big|\!\big|^2\leq\xi\,\big|\!\big|\,\widehat{\mathbf{x}}(t)\big|\!\big|^2$~~\textsf{or}~~ $t>T_\textsc{max}$} 
  \end{algorithmic}
  \caption{Sum-Product HyGAMP}
  \label{algo:gamp}
\end{algorithm*} 
 
 \indent In the sequel, we provide more details about HyGAMP alone which runs according to the algorithmic steps provided in Algorithm \ref{algo:gamp}. In our description, we assume that the hyperparameters $\sigma_x$ and $\sigma_w^2$ to be perfectly known to the receiver. Later on, we will explain how to also learn these two parameters from the data using the EM algorithm. For ease of exposition, the vector ${\mathbf{x}}_l$ to be reconstructed, in slot $l$, will be generically denoted as $\mathbf{x}$ since HyGAMP will be executed in each slot separately (same thing for $\mathbf{y}_l$ and all other quantities that depend on the slot index $l$). The  underlying block-sparse vector, $\mathbf{x}$, consists of $2^J$ blocks each of which consisting of $2M_r$ components, i.e., 
\begin{eqnarray}
\mathbf{x}~ \triangleq~ [ \mathbf{x}_1^{\textsf{T}},\mathbf{x}_2^{\textsf{T}},\ldots,\mathbf{x}_{2^J}^{\textsf{T}}]^\textsf{T},
\end{eqnarray}
with
 \begin{eqnarray}
 \mathbf{x}_j~ \triangleq~ [x_{1,j},x_{2,j},\ldots,x_{2M_r,j}]^\textsf{T},
 \end{eqnarray}
 Similarly, the known sensing matrix, $\mathbf{A}$, in (\ref{CS_Slot_l}) is partitioned into the corresponding $2^J$ blocks as follows:  
 \begin{equation}
 \mathbf{A}~ =~ \Big[\mathbf{A}^{(1)},\mathbf{A}^{(2)},\ldots,\mathbf{A}^{(2^J)}\Big]~~~\textrm{with}~~\mathbf{A}^{(j)}\in\mathbb{R}^{M\times2M_r}~\forall j.
 \end{equation}
 Recall also that $M = \frac{2nM_r}{L} $ and $N = 2^J(2M_r)$ denote the number of rows and columns in $\mathbf{A}$, respectively. 
 
\begin{figure*}[h!]
\hskip -0.1cm
    \includegraphics[scale=0.56]{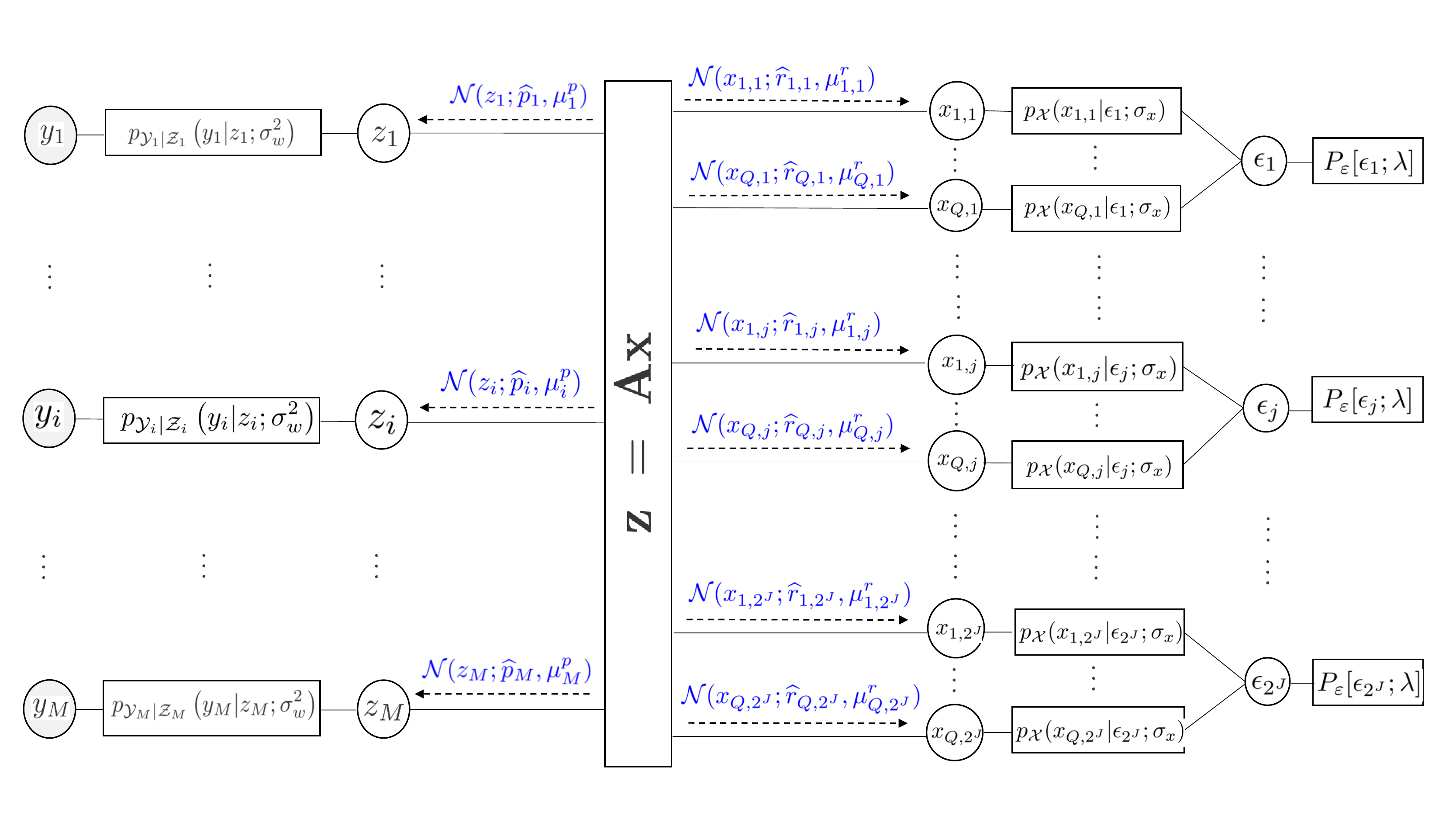}
    \caption{ Factor graph associated to the model in (\ref{CS_Slot_l}): $Q=2M_r$, $P_{\varepsilon}[\epsilon_j;\lambda]$ is the prior on $\epsilon_j$ given in (\ref{prior_epsilon}), $p_{\mathcal{X}}(x_{q,j}|\epsilon_j;\sigma_x)$ is the Bernoulli-Laplacian prior on $x_{q,j}$ given in (\ref{HyGamp:mixture_prior}).} 
    \label{factor_graph} 
\end{figure*}

 HyGAMP passes messages along the edges of the factor graph pertaining to  the model in (\ref{CS_Slot_l}) which is depicted in Fig. \ref{factor_graph}. There, the components of each block $\mathbf{x}_j$ are  connected to the same latent variable $\varepsilon_j$. The latter sends its belief (updated in line 20 of Algorithm 1) about each component of the block, being zero or non-zero. This updated belief is based on the information harvested from the other components of the same block (line 19 of \textbf{Algorithm 1}). In Fig. \ref{factor_graph}, the Gaussian messages  that are broadcast from the linear mix, $\mathbf{z}\,=\,\mathbf{A}\mathbf{x}$, to the variables nodes, $x_{q,j}$ and $z_i$, are highlighted in blue color. Their means and variances $\widehat{r}_{q,j}$, $\widehat{\mu}^r_{q,j}$, $\widehat{p}_{i}$, and $\widehat{\mu}^p_{i}$ are updated in lines
 9, 10, 17, and 18
 of Algorithm 1. The estimates of the unknown components in the group sparse vector are updated through the MMSE denoising step in line \ref{algo:mean_xn} and the associated variances are updated in line \ref{algo:var_xn}.
As a starting point, we initialize $\widehat{r}_{q,j}$ and $\mu_{q,j}^r$ $\forall q,j$ to $0$ and $1$, respectively. The LLRs and  sparsity-level  are initialized as in lines \ref{algo:LLR-init} and \ref{algo:sparsity-init}, respectively. As will be explained later, some of the updates must be derived based on the particular choice of the common prior which is  Bernoulli-Laplacian in this paper, i.e.:
\begin{eqnarray}\label{HyGamp:mixture_prior}
p_{\mathcal{X}}(x_{q,j}|\epsilon_j;\sigma_x)~=~(1-\epsilon_j)\delta(x_{q,j})~+~\epsilon_j \mathcal{L}(x_{q,j};\sigma_x),
\end{eqnarray}
\noindent where $\mathcal{L}(x;\sigma_x)$ is given in (\ref{laplacian_prior}) and $\delta(x)$ is the Dirac delta distribution. The updates for $\widehat{z}_i^0$ and $\mu_i^z$ (in lines \ref{algo:mean_zm} and \ref{algo:var_zm}) hold irrespectively of the prior since they depend only on the output distribution, namely,
\begin{eqnarray}
p_{\bm{\mathcal{Y}} |\bm{\mathcal{Z}}}\left(\mathbf{y} | \mathbf{z} ; \sigma_{w}^{2}\right)~=~\prod_{i=1}^{M} p_{\mathcal{Y}_{i} | \mathcal{Z}_{i}}\left(y_{i} | z_{i} ; \sigma_{w}^{2}\right),
\end{eqnarray}
in which $\mathbf{z} = \mathbf{A}\mathbf{x}$. Due to the AWGN channel assumption, our output distribution is Gaussian and we have the following updates readily available from \cite{GAMP}:
\begin{eqnarray}
\widehat{z}_{i}^{0}(t)~=~\frac{\mu_{i}^{p} y_{i}+\sigma_{w}^{2} \widehat{p}_{i}}{\mu_{i}^{p}+\sigma_{w}^{2}}
\end{eqnarray}
\begin{eqnarray}
\mu_{i}^{z}(t)~=~\frac{\mu_{i}^{p} \sigma_{w}^{2}}{\mu_{i}^{p}+\sigma_{w}^{2}}.
\end{eqnarray}
The updates in lines \ref{algo:mean_xn} and \ref{algo:var_xn}, however, depend on the particular choice of the prior and as such need to be expressed as function of  the other outputs of HyGAMP. In this paper, we only provide the final expressions of the required updates under the Bernoulli-Laplacian prior given in  (\ref{HyGamp:mixture_prior}). We omit the derivation details for sake of briefness since they are based on some equivalent algebraic manipulations as recently done in \cite{bellili2019generalized} in the absence of group sparsity. For notational convenience, we also introduce the following intermediate quantities that are needed to express the required updates, for $q=1,\ldots,2M_r$ and $j=1,\ldots,2^J$ (some variables are defined in Algorithm 1): 
\begin{eqnarray}
\theta_{q,j}&\triangleq&2 \sigma_x \,\frac{\left(1-\widehat{\rho}_{q,j}\right)}{\widehat{\rho}_{q,j} \sqrt{2 \pi \mu_{q,j}^{r}}}\\
\label{alpha-}
\alpha_{q,j}^{-}&\triangleq&
-\,\frac{\widehat{r}_{q,j}}{\sigma_x}~-~\frac{\mu_{q,j}^{r}}{2 \sigma_x^{2}},\\
\label{alpha+}\alpha_{q,j}^{+}&\triangleq&
\frac{\widehat{r}_{q,j}}{\sigma_x}~-~\frac{\mu_{q,j}^{r}}{2 \sigma_x^{2}}.
\end{eqnarray}
\begin{eqnarray}
\label{gamma-}\gamma_{q,j}^{-}&\triangleq&
\widehat{r}_{q,j}~+~\frac{\mu_{q,j}^{r}}{\sigma_x},\\
\label{gamma+}
\gamma_{q,j}^{+}&\triangleq&
\widehat{r}_{q,j}~-~\frac{\mu_{q,j}^{r}}{\sigma_x}.
\end{eqnarray}

It is worth mentioning here that those  quantities depend on the unknown parameter $\sigma_x$ of the Laplacian distribution. Therefore, on top of being updated by HyGAMP, these $\sigma_x-$dependent quantities must also be updated locally by the nested EM algorithm that learns the unknown parameter $\sigma_x$ itself. We also define the following two intermediate $\sigma_x-$independent quantities:
\begin{eqnarray}
\nu^{+}_{q,j}&\triangleq& Q\left(-\frac{\gamma_{q,j}^{+}}{\sqrt{\mu_{q,j}^{r}}}\right) e^{ \frac{\left(\gamma_{q,j}^{+}\right)^{2}}{2 \mu_{q,j}^{r}}},\\
\nu^{-}_{q,j}&\triangleq& Q\left(\frac{\gamma_{q,j}^{-}}{\sqrt{\mu_{q,j}^{r}}}\right) e^{\frac{\left(\gamma_{q,j}^{-}\right)^{2}}{2 \mu_{q,j}^{r}}},
\end{eqnarray}
in which $Q(.)$ is the standard Q-function, i.e., the tail of the normal distribution:
\begin{eqnarray}
Q(x)~=~\frac{1}{\sqrt{2\pi}}\int_{x}^{+\infty}\mathlarger{e^{\frac{-t^2}{2}}}dt.
\end{eqnarray}
Using the above notations, we establish the closed-form expressions\footnote{For full derivation details, see the most recent Arxiv version.} for  $\widehat{x}_{q,j}(t)$ required in line \ref{algo:mean_xn} of Algorithm 1 as given in (\ref{posterior_mean}) displayed on the top of the next page. For ease of notation, we drop the iteration index, $t$,  for all the statistical quantities updated by HyGAMP. The reader is referred to Algorithm 1 to keep track of the correct iteration count. The posterior variance, $\mu_{q,j}^{x}$,  required in  line \ref{algo:var_xn} is given by:
\begin{eqnarray}\label{posterior_variance}
\mu^x_{q,j}(t) &=&\sigma_{\mathcal{X}_{q,j}}^2\!(t)~-~\widehat{x}_{q,j}(t)^2,
\end{eqnarray}
wherein $\sigma_{\mathcal{X}_{q,j}}^2\!(t)$  is defined as follows:
\begin{eqnarray} 
\widetilde\sigma_{\mathcal{X}_{q,j}}^2\!(t)&&~\nonumber\\
&&\!\!\!\!\!\!\!\!\!\!\!\!\!\!\!\!\!\!\!\!\!\!\!\!\triangleq~\mathbb{E}_{\mathcal{X}_{q,j}|\bm{\mathcal{Y}}}\Big\{x_{q,j}^2|\mathbf{y};\widehat{r}_{q,j}(t-1),\mu^r_{q,j}(t-1),\widehat{\rho}_{q,j}(t),\sigma_x\Big\},\nonumber
\end{eqnarray}
and its closed-form expression is given by (\ref{posterior_variance}).
\begin{figure*} 
\begin{eqnarray}\label{posterior_mean}
\!\!\!\!\!\!\!\!\!\!\!\!\!\!\widehat{x}_{q,j}&=&\left(\frac{1}{\gamma_{q,j}^{-}+\gamma_{q,j}^{+} \frac{\nu^{+}_{q,j}}{\nu^{-}_{q,j}}}~+~\frac{1}{\gamma_{q,j}^{+}+\gamma_{q,j}^{-} \frac{\nu^{-}_{q,j}}{\nu^{+}_{q,j}}}~+~\frac{\theta_{q,j}}{\gamma_{q,j}^{+} \nu^{+}_{q,j}+\gamma_{q,j}^{-} \nu^{-}_{q,j}}\right)^{-1}.\\\nonumber\!\!\!\!\!\!\!\!\!\!\!\!\!\!&&\\
\!\!\!\!\!\!\!\!\!\!\!\!\!\!\label{posterior_variance}\frac{1}{\widetilde\sigma_{\mathcal{X}_{q,j}}^2}&=&
\left(\left[{(\gamma_{q,j}^{-})}^{2}+\mu_{q,j}^{r}\right]~+~\Big[(\gamma_{q,j}^{+})^{2}+\mu_{q,j}^{r}\Big] \frac{\nu^{+}_{q,j}}{\nu^{-}_{q,j}}~-~\frac{2(\mu_{q,j}^{r})^2}{\sigma_x\sqrt{2\pi\mu_{q,j}^{r}}\nu^{-}_{q,j}}\right)^{-1}\nonumber\\&& 
~~~~~~~~~~~~~~~~~~~+\left(\Big[(\gamma_{q,j}^{+})^{2}+\mu_{q,j}^{r}\Big]~+~\Big[(\gamma_{q,j}^{-})^{2}+\mu_{q,j}^{r}\Big]\frac{\nu^{-}_{q,j}}{\nu^{+}_{q,j}}~ -~ \frac{2(\mu_{q,j}^{r})^2}{\sigma_x\sqrt{2\pi\mu_{q,j}^{r}}\nu^{+}_{q,j}}\right)^{-1} \nonumber\\&&
~~~~~~~~~~~~~~~~~~~~~~~~~~~~~~~~~~~~~~~~+~\theta_{q,j}\left(\Big[(\gamma_{q,j}^{+})^{2}+\mu_{q,j}^{r}\Big] \nu^{+}_{q,j} ~+~ \Big[(\gamma_{q,j}^{-})^{2}+\mu_{q,j}^{r}\Big] \nu^{-}_{q,j}~-~\frac{2(\mu_{q,j}^{r})^2}{\sigma_x\sqrt{2\pi\mu_{q,j}^{r}}}\right)^{-1}\!\!\!\!\!\!.
\end{eqnarray}
\end{figure*}
\begin{figure*}
\vspace{-0.75cm} 
\begin{tabular}{  r r r r r r r r r r r r r r r r r r r r r r r r r r r r  r r r r r r r r r r r r r r r r r r   }
&~&~&~&~&~&~&~&~&~&~&~&~&~&~&~&~&~&~&~&~&~&~&~&~&~&~ &~&~&~&~&~&~&~&~&~&~&~&~&~&~&~&~\\
  \hline
\end{tabular}
\end{figure*}
The closed-form expression for the LLR update in line \ref{algo:LLR_reverse} of Algorithm 1 was also established as follows:
\begin{eqnarray}\label{LLR_update}
\textrm{LLR}_{q\rightarrow j }\,~=\,~\log \left(\frac{\sqrt{2 \pi \mu_{q,j}^{r}}}{2 \sigma_x}\big[\nu^{+}_{q,j}+\nu^{-}_{q,j}\big]\right).
\end{eqnarray}
  We also resort to the maximum likelihood (ML) concept in order to estimate  the unknown hyperparameters $\sigma_x$ and $\sigma_w^2$. More specifically, the  ML estimate of the noise variance is given by:
\begin{eqnarray}
\widehat{\sigma}_{w}^{2}~=~\frac{1}{M} \textstyle\sum_{i=1}^{M}\left(y_{i}-\widehat{z}_{i}\right)^{2}+\mu_{i}^{z},
\end{eqnarray}
where $\widehat{z}_{i}\triangleq (\mathbf{A}\widehat{\mathbf{x}})_i$. Unfortunately, 
 the ML estimate (MLE), $\widehat{\sigma}_x$, of $\sigma_x$, cannot be found in closed form and we use the EM algorithm instead to find the required MLE iteratively. Indeed, starting from some initial guess, $\widehat{\sigma}_{x;0}$,  we establish the $(d+1)$th MLE update as follows: 
\begin{eqnarray}
\widehat{\sigma}_{x;d+1} ~=~ \frac{1}{\sum_{q,j}\widehat{\rho}_{q,j}}\sum_{q,j}\frac{\widehat{\rho}_{q,j}\,\kappa_{q,j;d}}{2\widehat{\sigma}_{x;d}\,\psi_{q,j;d}}, 
\end{eqnarray}

in which the quantities $\psi_{q,j;d}$ and $\kappa_{q,j;d}$ are given by:
%
\begin{eqnarray}\label{b_hat_related_psi}
\!\!\!\!\!\!\!\!\!\!\!\!\!\!\psi_{q,j;d} &=&\frac{\widehat{\rho}_{q,j}}{2\sigma_{x;d}}\left[\mathlarger{e^{-\alpha^{-}_{q,j;d}}}Q\left(\frac{\gamma^{-}_{q,j;d}}{\sqrt{\mu^{r}_{q,j}}}\right)\right.\nonumber\\
&& ~~~~~~~~~~+~\left. \mathlarger{e^{-\alpha^{+}_{q,j;d}}}Q\left(-\frac{\gamma^{+}_{q,j;d}}{\sqrt{\mu^{r}_{q,j}}}\right)\right] \nonumber\\
&&~~~~~~~~~~~~~~~~~~~~~~~~~~~+~(1-\widehat{\rho}_{q,j})\mathlarger{e^{-\frac{r^2_{q,j}}{2\mu^{r}_{q,j}}}},\\\nonumber\\
\!\!\!\!\!\!\!\!\!\!\!\!\! \label{b_hat_related_kappa}\kappa_{q,j;d} &=& \gamma^{+}_{q,j;d}\,\mathlarger{e^{-\alpha^{+}_{q,j;d}}}Q\left(\frac{\gamma^{+}_{q,j;d}}{\sqrt{\mu^{r}_{q,j}}}\right) \nonumber\\
&&~~~~~~~~~~-~ \gamma^{-}_{q,j;d}\,\mathlarger{e^{-\alpha^{-}_{q,j;d}}}Q\left(\frac{\gamma^{-}_{q,j;d}}{\sqrt{\mu^{r}_{q,j}}}\right) \nonumber\\
&&~~~~~~~~~~~~~~~~~~~~~~~~~~~~+~  \frac{2\mu^{r}_{q,j}}{\sqrt{2\pi\mu^{r}_{q,j}}}\mathlarger{e^{-\frac{r^2_{q,j}}{2\mu^{r}_{q,j}}}}. 
\end{eqnarray}
Note here that $\gamma^{+}_{q,j;d}$, $\gamma^{-}_{q,j;d}$, $\alpha^{+}_{q,j}$, and $\alpha^{-}_{q,j;d}$ involved in (\ref{b_hat_related_psi})-(\ref{b_hat_related_kappa}) are also expressed as in (\ref{alpha-})-(\ref{gamma+}), except the fact that $\sigma_x$ is now replaced by $\widehat{\sigma}_{x;d}$.

\subsection{Constrained Clustering-Based Stitching Procedure}
In this section, we focus on the problem of clustering the reconstructed channels from all the slots to obtain one cluster per user. By doing so, it will be easy to cluster (i.e., stitch) the slot-wise decoded sequences of all users so as to recover their  transmitted messages/packets. To that end, we first estimate the large-scale fading coefficients from the outputs of HyGAMP as follows: 
\begin{eqnarray}\label{large_scale_estimate}
\widehat{g}_{k,l} ~=~ \frac{1}{M_r}\big\|\widehat{\mathbf{h}}_{k,l}\big\|_2^2,
\end{eqnarray}
where $\widehat{\mathbf{h}}_{k,l}$ is the $k$th reconstructed channel in slot $l$. The estimates of the different large-scale fading coefficients are required to re-scale the reconstructed channels before clustering. This is in order to avoid, for instance, having the channels of the cell-edge users clustered together due to their strong pathloss attenuation.  To that end, we divide each $\widehat{\mathbf{h}}_{k,l}$ in (\ref{reconstructed_channels}) by the associated $\sqrt{\widehat{g}_{k,l}}$ in (\ref{large_scale_estimate}) but keep using the same symbols, $\widehat{\mathbf{h}}_{k,l}$, for notational convenience. 

We can then visualize  (\ref{reconstructed_channels}) --- after  normalization --- as one whole set of $K_aL$ data points in $\mathbb{R}^{2M_r}$:
\begin{eqnarray}
\mathcal{H}~ =~ \{\mathbf{\widehat{h}}_{k,l}\mid k=1,\ldots,K_a,~~l=1,\ldots,L\}.
\end{eqnarray}
which  gathers all the reconstructed small-scale fading coefficients pertaining to all $K_a$ active users and all $L$ slots. Since the small scale-fading coefficients of each user  are assumed to be Gaussian distributed, we propose to fit a Gaussian mixture distribution to the entire data set, $\mathcal{H}$ , and use the EM algorithm to estimate the parameters of the involved mixture densities along with the mixing coefficients. 

The rationale behind the use of clustering is our prior knowledge about the nature of the data set $\mathcal{H}$. Indeed, we know that there are $K_a$ users whose channels remain constant over all the slots. Therefore, each user contributes exactly $L$ data points in $\mathcal{H}$ which are noisy estimates of its true channel vector. Our goal is  hence to cluster the whole data set into $K_a$ different clusters, each of which having exactly $L$ vectors. To do so, we denote the total number of data points in $\mathcal{H}$ by $N_{\textrm{tot}} \triangleq {K}_aL$ and assume that each data point is an independent realization of a Gaussian-mixture distribution with ${K}_a$ components:
\begin{eqnarray}\label{Gaussian_mixture}
p_{\bm{\mathcal{\widehat{H}}}}(\mathbf{\widehat{h}};\bm{\pi},\bm{\mu},\bm{\Sigma})~=~ \sum_{k=1}^{{K}_a}{\pi}_k\,\mathcal{N}(\mathbf{\widehat{h}};\bm{{\mu}}_k,\bm{{\Sigma}}_k).
\end{eqnarray}
Here, $\bm{\pi}\,\triangleq\,[\pi_1,...,\pi_{K_a}]^\textsf{T}$ are the mixing coefficients, $\bm{\mu} \triangleq [\bm{\mu}_1,...,\bm{\mu}_{K_a}]^\textsf{T}$ are the clusters' means, and $\bm{\Sigma} \triangleq [\bm{\Sigma}_1,...,\bm{\Sigma}_{K_a}]^\textsf{T}$ are their covariance matrices.

The assumption we make here is justified by the fact that the residual reconstruction noise of AMP-like algorithms (including HyGAMP) is Gaussian-distributed. Notice that in (\ref{Gaussian_mixture}) we considered a mixture of $K_a$ components, which amounts to assigning a Gaussian distribution to each active user. We now turn our attention to finding the likelihood function of all the unknown parameters\footnote{Note here that we refer to each $\bm{\mu}_k$ and $\bm{\Sigma_k}$ as parameters although strictly speaking they are vectors and matrices of unknown parameters.}, $\{\pi_k,\bm{{\mu}}_k,\bm{{\Sigma}}_k\}_{k=1}^{K_a}$, involved in (\ref{Gaussian_mixture}). To that end, we use $\widehat{\mathbf{h}}_n$ to denote a generic data point in $\mathcal{H}$, i.e.:
\begin{eqnarray}
\mathcal{H} &=& \left\{\mathbf{\widehat{h}}_{k,l}\mid  k=1,\ldots,K_a,~~l=1,\ldots,L\right\},\\
&=&\Big\{\widehat{\mathbf{h}}_n\,|\,n=1,\ldots,N_{\textrm{tot}}\Big\}.
\end{eqnarray}
Owing to the i.i.d assumption on the data, the associated likelihood function factorizes as follows:
\begin{equation}\label{factorized_mixture}
  p_{\bm{\mathcal{\widehat{H}}}_1,\ldots,\bm{\mathcal{\widehat{H}}}_{N_{\textrm{tot}}}}(\mathbf{\widehat{h}}_1,\ldots,\mathbf{\widehat{h}}_{N_{\textrm{tot}}};\bm{\pi},\bm{\mu},\bm{\Sigma})~=~ \prod_{n=1}^{{N_{\textrm{tot}}}}p_{\bm{\mathcal{\widehat{H}}}}(\mathbf{\widehat{h}}_n;\bm{\pi},\bm{\mu},\bm{\Sigma}).
\end{equation}
Taking the logarithm of (\ref{factorized_mixture}) yields the following log-likelihood function (LLF):
 \begin{eqnarray}
 \mathfrak{L}\big(\bm{\pi},\bm{\mu},\bm{\Sigma}\big)&\triangleq\ln p_{\bm{\mathcal{\widehat{H}}}_1,\ldots,\bm{\mathcal{\widehat{H}}}_{N_{\textrm{tot}}}}(\mathbf{\widehat{h}}_1,\ldots,\mathbf{\widehat{h}}_{N_{\textrm{tot}}};\bm{\pi},\bm{\mu},\bm{\Sigma}),\nonumber\\
 &=\sum_{n=1}^{{N_{\textrm{tot}}}}\ln\left( \sum_{k=1}^{{K}_a}{\pi}_k\,\mathcal{N}(\mathbf{\widehat{h}}_n;\bm{\mu}_k,\mathbf{{\Sigma}}_k)\right).
\end{eqnarray}
 Our task is to then maximize the LLF with respect to the unknown parameters, i.e.:
 \begin{eqnarray}
 \argmax_{{\pi}_k,\bm{\mu}_k,\mathbf{\Sigma}_k}\sum_{n=1}^{{N_{\textrm{tot}}}}\ln\left( \sum_{k=1}^{{K}_a}{\pi}_k\,\mathcal{N}(\mathbf{\widehat{h}}_n;\bm{\mu}_k,\mathbf{{\Sigma}}_k)\right).
\end{eqnarray}

Unfortunately, it is not possible to obtain a closed-form solution to the above optimization problem. Yet, the EM algorithm can again be used to iteratively update the ML estimates of the underlying parameters. In the sequel, we will provide the resulting updates, and we refer the reader to Chap. 9 of  \cite{bishop2006pattern} for more details.

We initialize the means, $\{\bm{\mu}_k\}_{k=1}^{K_a}$, using \textit{k-means++} sample which outputs the $K_a$ centroids of the original data set. Covariance matrices, $\{\bm{\Sigma}_k\}_{k=1}^{K_a}$, are initialized to be diagonal with the diagonal elements chosen as the means of the rows of the data set, and mixing coefficients $\{\pi_k\}_{k=1}^{K_a}$ are initialized uniformly as $1/K_a$. Then, the three steps needed to learn the parameters of the above Gaussian mixture model are as follows:

\begin{itemize}
\item \textbf{Expectation step (\textsc{E-STEP})}
\begin{eqnarray}
\Pr(\mathbf{\widehat{h}}_n\in \textrm{cluster}~ k)~=~ \frac{\pi_k\,\mathcal{N}(\mathbf{\widehat{h}}_n;\bm{\mu}_k,\mathbf{\Sigma}_k)}{\sum_{k=1}^{{K}_a}{\pi}_k\,\mathcal{N}(\mathbf{\widehat{h}}_n;\bm{{\mu}}_k,\mathbf{{\Sigma}}_k)}.
\end{eqnarray}
\item \textbf{Maximization step (\textsc{M-STEP})}
\begin{eqnarray}
\!\!\!\!\!\!\!\!\!\!\!\!\!\!\!\!\bm{\mu}_k^{\textrm{new}} &=& \frac{1}{N_k}\sum_{n=1}^{{N_{\textrm{tot}}}}\Pr(\mathbf{\widehat{h}}_n\in \textrm{cluster}~ k)\mathbf{\widehat{h}}_n,\\
\!\!\!\!\!\!\!\!\!\!\!\!\!\!\!\!\mathbf{\Sigma}_k^{\textrm{new}} 
&=& \frac{1}{N_k}\sum_{n=1}^{{N_{\textrm{tot}}}}\Pr(\mathbf{\widehat{h}}_n\in \textrm{cluster}~ k)\\
\!\!\!\!\!\!\!\!\!\!\!\!\!\!\!\!&&~~~~~~~~~~~~~\times(\mathbf{\widehat{h}}_n-\bm{{\mu}}_k^{\textrm{new}})(\mathbf{\widehat{h}}_n-\bm{{\mu}}_k^{\textrm{new}})^\textsf{T},\\
\!\!\!\!\!\!\!\!\!\!\!\!\!\!\!\!{N}_k &=&\sum_{n=1}^{{N_{\textrm{tot}}}}\Pr(\mathbf{\widehat{h}}_n\in \textrm{cluster}~ k),\\
\!\!\!\!\!\!\!\!\!\!\!\!\!\!\!\!\pi_k^{\textrm{new}} &=& \frac{N_k}{{N_{\textrm{tot}}}}.
\end{eqnarray}
\item \textbf{Evaluation step (\textsc{Eval-STEP})}
 \begin{equation}
  \!\!\!\!\!\!\!\!\mathfrak{L}\big(\bm{\pi^\textrm{new}},\bm{\mu^\textrm{new}},\bm{\Sigma^\textrm{new}}\big)= \sum_{n=1}^{{N_{\textrm{tot}}}}\ln\left( \sum_{k=1}^{{K}_a}{\pi}_k^\textrm{new}\mathcal{N}(\mathbf{\widehat{h}}_n;\bm{{\mu}}_k^\textrm{new},\mathbf{{\Sigma}}_k^\textrm{new})\!\right)\!\!.
\end{equation}
\end{itemize}

In the E-\textsc{STEP}, we compute the probability of having a particular data point belong to each of the $K_a$ users. In the M-\textsc{STEP}, we update the means, covariance matrices, and the mixing coefficients for each of the clusters. We further need to evaluate the LLF at each iteration to check the convergence of the EM-based algorithm, hence the Eval-STEP. Recall, however, that we are actually dealing with a constrained clustering problem since it is mandatory to enforce the following two intuitive constraints:
\begin{itemize}
\item \textbf{Constraint 1\textsc{:}}\label{constraint_1}
 Channels from the same slot cannot be assigned to the same user, 
 \item \textbf{Constraint 2\textsc{:}}
 Users/clusters should have exactly $L$  channels/data points.
\end{itemize}

At convergence, the EM algorithm returns a matrix, $\mathbf{P}$, of posterior membership probabilities, i.e., whose $(n,k)$th entry is $\mathbf{P}_{nk} = \Pr(\widehat{\mathbf{h}}_n\in \textrm{cluster}~ k)$. Since the EM solves an unconstrained clustering problem, relying directly on $\mathbf{P}$ would result in having two channels reconstructed from the same slot being  clustered together, thereby violating ``constraint 1'' and/or ``constraint 2''.   In what follows, we will still make use of $\mathbf{P}$ in order to find the best possible assignment of the ${N_{\textrm{tot}}}$ reconstructed channels to the $K_a$ users (i.e., the one that minimizes the probability of error) while satisfying the two constraints mentioned above.\\ 
To enforce ``constraint 2'', we begin by partitioning $\mathbf{P}$ into $L$ equal-size and consecutive blocks, i.e.,  $K_a\times K_a$ matrices $\{\mathbf{P}^{(l)}\}_{l=1}^{L}$, as follows:
\begin{eqnarray}
\mathbf{P}
~=~\begin{bmatrix}
\mathbf{P}^{(1)}\\
\cdots\cdots\cdots\\
\vdots\\
\cdots\cdots\cdots\\\mathbf{P}^{(L)}\\
\end{bmatrix}.
\end{eqnarray}
Then, since each $k$th row in $\mathbf{P}^{(l)}$ sums to one, it can be regarded as a distribution of some categorical random  variable, $\mathcal{V}_{k,l}$, that can take on one of $K_a$ possible mutually exclusive states. For convenience, we represent these categorical random variables by $1$-of-$K_a$ binary coding scheme. That is, each $\mathcal{V}_{k,l}$ is represented by a $K_a$-dimensional
vector $\mathbf{v}_{k,l}$ which takes values in $\{\mathbf{e}_1, \mathbf{e}_2,..., \mathbf{e}_{K_a}\}$ where $\mathbf{e}_i=[0,...,1,...,0]^{\textsf{T}}$  has a single $1$ located  at position $i$. We also denote the set of all $K_a\times K_a$ permutation matrices by $\mathcal{P}$.



 We enforce  ``constraint 1'' by using the following posterior joint distribution on $\{\mathcal{V}_{k,l}\}_{k=1}^{K_a}$ in each $l$th slot:
\begin{equation}\label{joint_of_Z}
p_{\mathcal{V}_{1,l},\ldots,\mathcal{V}_{K_a,l}}(\mathbf{v}_{1,l},\ldots,\mathbf{v}_{K_a,l})~\propto~ \mathbb{I}(\mathbf{V}_l\in \mathcal{P})\prod_{k=1}^{K_a}p_{\mathcal{V}_{k,l}}(\mathbf{v}_{k,l}),
\end{equation}
where $\mathbb{I}(.)$ is the indicator function and $\mathbf{V}_l \triangleq [\mathbf{v}_{1,l},\ldots,\mathbf{v}_{K_a,l}]^{\textsf{T}}$. Moreover, it is clear that any categorical distribution with $K_a$ atoms can be parametrized in the following way:
\begin{eqnarray}\label{marginal_of_Z}
p_{\mathcal{V}_{k,l}}(\mathbf{v}_{k,l})~=~ \exp\left\{\sum_{k'=1}^{K_a}\alpha_{k,k'}^{(l)}\mathbb{I}(\mathbf{v}_{k,l}=\mathbf{e}_{k'})\right\},
\end{eqnarray}
in which
\begin{eqnarray}
\alpha_{k,k'}^{(l)}~\,=~\, \log p_{\mathcal{V}_{k,l}}(\mathbf{v}_{k,l} = \mathbf{e}_{k'})~\,=~\, \log \mathbf{P}^{(l)}_{k,k'}.
\end{eqnarray} 
Since our optimality criteria is the largest-probability assignment, we need to maximize the distribution in (\ref{joint_of_Z}) which when combined with (\ref{marginal_of_Z}) yields:
\begin{eqnarray}\label{posterior_exp}
p_{\bm{\mathcal{V}}_l}(\mathbf{V}_l)&\!=\!&p_{\mathcal{V}_{1,l},\ldots,\mathcal{V}_{K_a,l}}(\mathbf{v}_{1,l},\ldots,\mathbf{v}_{K_a,l}),\nonumber\\
&\!\propto\!&\mathbb{I}(\mathbf{V}_l\in \mathcal{P})\prod_{k=1}^{K_a}\exp\left\{\sum_{k'=1}^{K_a}\alpha_{k,k'}^{(l)}\mathbb{I}(\mathbf{v}_{k,l}=\mathbf{e}_{k'})\right\},\nonumber\\
&\!=\!&\label{posterior_clustering_final}\mathbb{I}(\mathbf{V}_l\in \mathcal{P})\exp\left\{\sum_{k=1}^{K_a}\sum_{k'=1}^{K_a}\alpha_{k,k'}^{(l)}\mathbb{I}(\mathbf{v}_{k,l}=\mathbf{e}_{k'})\right\}.\nonumber\\
\end{eqnarray}
Now, finding the optimal assignment inside slot $l$, subject to constraint 1, amounts to  finding the optimal assignment   matrix, $\widehat{\mathbf{V}}_l$, that maximizes the constrained posterior joint distribution, $p_{\bm{\mathcal{V}}_l}(\mathbf{V}_l)$, established in (\ref{posterior_exp}), i.e.:
\begin{eqnarray}
\widehat{\mathbf{V}}_l ~=~ \argmax_{\mathbf{V}_l} p_{\bm{\mathcal{V}}_l}(\mathbf{V}_l). 
\end{eqnarray}
Owing to (\ref{posterior_exp}), it can be shown that finding $\widehat{\mathbf{V}}_l$ is equivalent  to solving the following constrained optimization problem:
\begin{eqnarray}
\argmax_{\mathbf{V}_l}\, \sum_{k=1}^{K_a}\sum_{k'=1}^{K_a}\alpha_{k,k'}^{(l)}\mathbb{I}(\mathbf{v}_{k,l}=\mathbf{e}_{k'})
\end{eqnarray}

\begin{eqnarray}\label{signum_function}
\textrm{subject to} \begin{cases}
~\sum_{k'=1}^{K_a}\mathbb{I}(\mathbf{v}_{k,l}=\mathbf{e}_{k'})=1 ~~~\textrm{for all}~ k\\
~ \sum_{k=1}^{K_a}\mathbb{I}(\mathbf{v}_{k,l}=\mathbf{e}_{k'})=1 ~~~\textrm{for all}~ k'.\\
\end{cases}
\end{eqnarray}
Note that the constraints in (\ref{signum_function}) enforce the solution to be a permutation matrix. This follows from the factor, $\mathbb{I}(\mathbf{V}_l\in \mathcal{P})$, in the posterior distribution established in  (\ref{posterior_exp}) which assigns zero probability to non-permutation matrices. 

This optimization problem 
can be solved in polynomial time by means of the Hungarian algorithm which has an overall complexity in the order of $\mathcal{O}(K_a^3)$. Stitching  is achieved by means of the optimal assignment matrices, $\{\widehat{\mathbf{V}}_l\}_{l=1}^{L}$, which are used to cluster the reconstructed sequences, thereby recovering the original transmitted messages.

\section{Simulation Results}\label{section_5}
\subsection{Simulation Parameters}
In this section, we assess the performance of the proposed URA scheme using exhaustive Monte-Carlo computer simulations. Our performance metric is the probability of error given in (\ref{wireless:error_prob}). We fix the number of information bits per user/packet to $B \,=\, 102$,  which are communicated over  $L \,=\, 6$ slots. This corresponds to $J \,=\, 17$ information  bits per slot.  We also fix the bandwidth to $W = 10$ MHz and the noise power to $P_{w} = 10^{-19.9}\times W$  [Watts].
  The path-loss  parameters in (\ref{wireless:large-scale_fading}) are set to $\alpha\,=\,-15.3$ dB and $\beta\,=\,3.76$. 
The users are assumed  to be uniformly scattered  on an annulus centered at the base station and with  inner and outer radiuses, $R_{in} = 5$ meters and $R_{out} = 1000$ meters, respectively. The distribution of each $k$th user random distance, $R_k$, from the base station is hence given by:
\begin{equation}
    \Pr(R_k < r_k) ~=~ \frac{r_k^2 - R_{in}^2}{R_{out}^2 - R_{in}^2}.
\end{equation}
  In the following, our baseline is the covariance-based scheme introduced recently in \cite{5G_2} which is simply referred to as CB-CS in this paper. For the CB-CS algorithm we fix the number of information bits per user/packet to $B = 104$ bits which are communicated over $L = 17$ slots. The parity bit allocation for the outer tree code was set to $p = [0,8,8,\ldots,14]$. We also use $J = 14$ coded bits per slot which leads to the total rate of the outer code $R_{out} = 0.437$.
  \subsection{Results}
\begin{figure}[!ht] 
    \vskip -0.5cm
  \hskip -0.4cm 
    \includegraphics[scale=0.8]{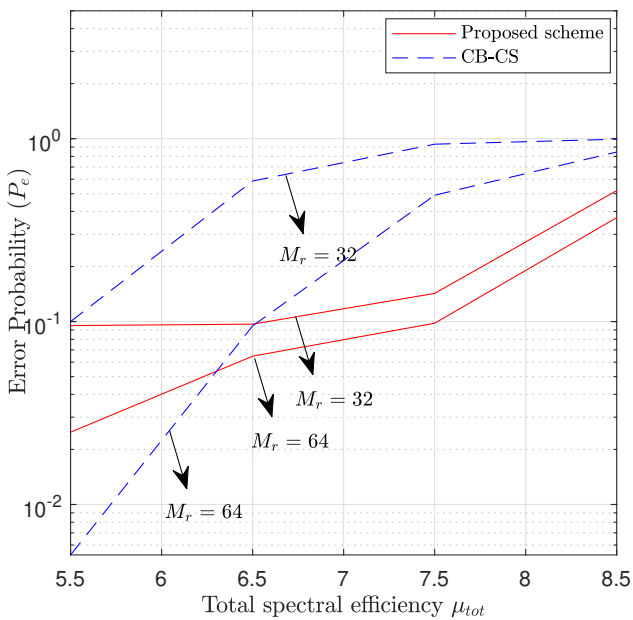}
    \caption{Performance of the proposed scheme as a function of the total spectral efficiency, $\mu_{tot}$,  and receive antenna elements, $M_r$, with a fixed number of active users $K_a\,=\,150$ and transmit power $P_t = 15$ dBm.}
    \label{fig:comparison of the schemes_Ptx_15}
\end{figure}

\begin{figure}[!ht]
  \vskip -0.5cm
  \hskip -0.4cm
    \includegraphics[scale=0.68]{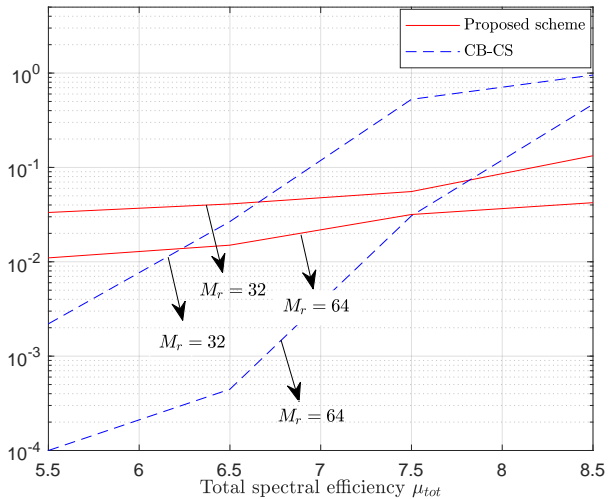}
    \caption{Performance of the proposed scheme as a function of the total spectral efficiency, $\mu_{tot}$,  and receive antenna elements, $M_r$, with a fixed number of active users $K_a\,=\,150$ and transmit power $P_t = 20$ dBm.}
    \label{fig:comparison of the schemes_Ptx_20} 
\end{figure}

  Figs. \ref{fig:comparison of the schemes_Ptx_15} and \ref{fig:comparison of the schemes_Ptx_20} depict the performance of both URA schemes as a function of the total spectral efficiency $\mu_{tot}\,=\,K_aB/n$. In Fig. \ref{fig:comparison of the schemes_Ptx_15}, we fix the transmit power to $P_t\,=\,15$ dBm for all the users and show two curves for two different number of antennas, namely $M_r = 32$ and $M_r = 64$. Similar setting is depicted in Fig. \ref{fig:comparison of the schemes_Ptx_20} except the transmit power is increased to $P_t\,=\,20$ dBm. The total number of users in both plots is fixed to $K_a = 150$. From the total spectral efficiency it is possible to calculate the blocklength required for both  CB-CS and the proposed URA scheme using $n = (B K_a)/\mu_{tot}$ and  the blocklength per slot $n_0 = n/L$. At the smallest spectral efficiency considered in Figs. \ref{fig:comparison of the schemes_Ptx_15} and \ref{fig:comparison of the schemes_Ptx_20} (i.e., $\mu_{tot} = 5.5$ bits/channel-use), the blocklength per-slot of the proposed scheme becomes $n_0 = 464$. This yields a sensing matrix $\mathbf{A}$ which has  $N_{\textrm{row}}\,=\,(2M_r)\times464$ rows and $N_{\textrm{col}}\,=\,(2M_r)\times2^{17}$ columns. As a matter of fact, when $M_r = 64$ we have $N_{\textrm{row}}\,\sim\,10^4$ and  $N_{\textrm{col}}\,\sim\,10^7$ which is a very large dimension for the multiple matrix-vector multiplications required inside HyGAMP. To alleviate this computational burden, we use a circulant Gaussian codebook $\widetilde{\mathbf{A}}$ so as to perform these multiplications via the fast Fourier transform (FFT) algorithm. Huge computational savings also follow from taking advantage of the inherent Kronecker structure in $\mathbf{A}$ involved in (\ref{sireless:system_model_vec}). \\\indent 

\begin{figure}[!thb]
  \hskip -0.5cm
    \centering
    \includegraphics[scale=0.8]{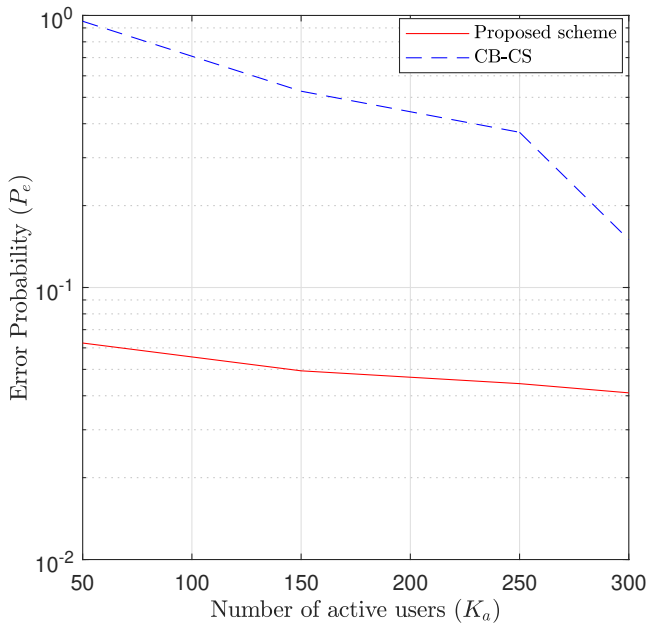}
    \caption{Performance of the proposed scheme as a function of the number of active users, $K_a$, with fixed total spectral efficiency, $\mu_{tot} = 7.5$ bits/channel-use, number of receive antennas, $M_r = 32$, and transmit power $P = 20$ dBm.} 
    \label{fig:total_spectral_efficiency}
\end{figure}

 As can be seen from Figs. \ref{fig:comparison of the schemes_Ptx_15} and \ref{fig:comparison of the schemes_Ptx_20},  the proposed scheme outperforms  CB-CS when the total spectral efficiency becomes large as is desirable in massive connectivity setups. The inefficiency of CB-CS stems from the fact that at high total spectral efficiency the number of active users exceeds both the number of antenna branches at the BS and the per-slot blocklength.  
   The proposed URA scheme is able to support high spectral efficiencies  by making use of the small-scale fading signatures of the different users to stitch sequences instead of relying on concatenated coding which reduces the effective data rate. In fact, as the number of antennas increases, the users' channels become almost orthogonal and users can be easily separated in the spatial domain due to the higher spatial resolution and the channel hardening effect, which is one of the blessings of massive MIMO. \\ 
    It is worth mentioning, however, that by eliminating the outer code one expects a net gain of $1/R_{\textrm{outer}}$ in spectral efficiency under perfect CSI. In this respect, we emphasize the fact that at $P_t = 20$ dBm and small spectral efficiencies it was found that HyGAMP provides quasi-perfect CSI. Therefore, the error floor observed at the low-end spectral efficiency in Figs. \ref{fig:comparison of the schemes_Ptx_15} and \ref{fig:comparison of the schemes_Ptx_20} is due to the inefficiencies  of EM-based clustering. 
   Using the Bayes optimal clustering decoder \cite{lesieur2016phase} instead of EM is hence an interesting research topic which we plan to investigate in the near future.\\
 Now, we turn the tables and fix the total spectral efficiency to $\mu_{\textrm{tot}}= 7.5$  bits/channel-use while varying the number of active users from $K_a=50$ to $K_a=300$. The performance of both URA schemes is depicted in Fig. \ref{fig:total_spectral_efficiency}. There, the number of receive antennas and the transmit power were fixed to $M_r=32$ and $P_t = 20$ dBm, respectively. At such high total spectral efficiency, the proposed scheme achieves a decoding error probability $P_e\sim 10^{-2}$ by using 32 antennas only. 
The net performance gains brought by the proposed scheme are of course a direct consequence of removing the concatenated coding thereby providing the HyGAMP CS decoder with more observations to reconstruct the channels as apposed to the inner CS decoder of the CB-CS scheme.

  
\begin{figure}[!thb]
  \hskip -0.5cm
    \centering
    \includegraphics[scale=0.80]{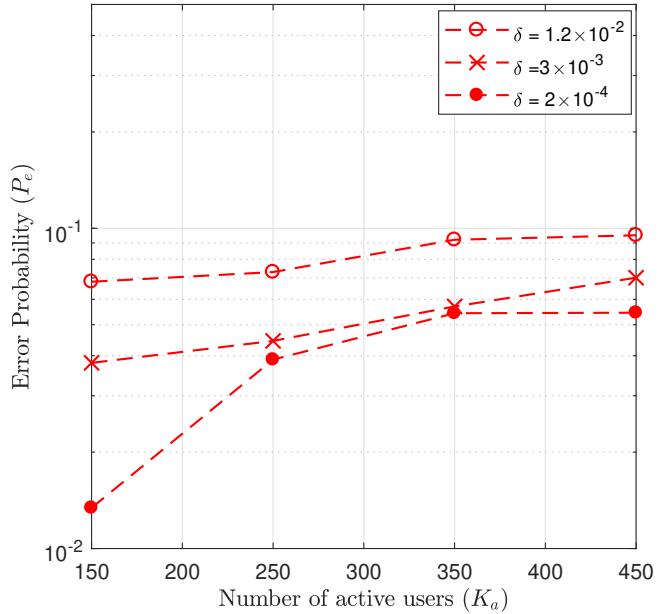}
    \caption{Performance of the proposed scheme under inter-slot channel time variations as a function of the number of active users, $K_a$, with fixed per-user spectral efficiency, $\mu = 0.015$  bits/channel-use/user, number of receive antennas, $M_r = 32$, and transmit power $P_t = 15$ dBm.} 
    \label{fig:time_variation}
\end{figure}

  In Fig. \ref{fig:time_variation}, we asses the performance of the proposed scheme as a function of the number of active users, $K_a$, with channel time variations across slots (i.e., with slot-wise block-fading only).  The decorrelation of the channel between two consecutive slots is denoted by $\delta$. Consider a rich scattering environment in which the channel correlation, at dicrete time lag $k$, follows the  Clarke-Jakes' model: 
  \begin{equation}\label{Clarke}
    R_h[k] = J_0\left(2\pi f_c\frac{kv}{Wc}\right) = \sqrt{(1-\delta)},
\end{equation}
wherein $J_0(.)$  is the zero-order Bessel function of the first kind, $f_c = 2$ GHz is the carrier frequency, $v$ is the relative velocity between the receiver and the transmitter, and $c=3\times10^8$ m/s is the speed of light.  In Fig. \ref{fig:time_variation} we investigate three different mobility regimes at per-user spectral efficiency $\mu\,=\,0.015$ bits/user/channel-use. This corresponds to $k = 1133$ which can be used in (\ref{Clarke}) to find the value of $\delta$ associated to each relative velocity $v$ as follows:  
\begin{enumerate}
\item Typical pedestrian scenario ($v = 5$ km/h) for which  $\delta = 0.00002$, 
\item Typical urban scenario ($v = 60$ km/h) for which  $\delta = 0.00312$,
\item Typical vehicular scenario ($v = 120$ km/h) for which  $\delta = 0.01244$.
\end{enumerate}
The  simulation results shown in Fig. \ref{fig:time_variation}  reveal that the performance of the proposed algorithm is still acceptable even under the high-speed scenario (i.e., up to $v = 120$ km/h). This is because the EM algorithm is able to capture the inter-slot/intra-cluster time correlations through the updated covariance matrices.   

\section{Conclusion}\label{section_6}
We have introduced a new algorithmic solution to the unsourced random access problem that is also based on slotted transmissions. As opposed to all existing works, however, the proposed scheme relies purely on the rich spatial dimensionality offered by large-scale antenna arrays instead of coding-based coupling for sequence stitching purposes. HyGAMP CS recovery algorithm has been used to reconstruct the users channels and decode the sequences on a per-slot basis. Afterwards, the EM framework together with the Hungarian algorithm  have been used to solve the underlying constrained clustering/stitching problem. The performance of the proposed approach has been compared  to the only existing URA algorithm, in the open literature. The proposed scheme provides  performance enhancements in a high spectral efficiency regime.
There are many possible avenues for future work. The two-step procedure of channel estimation and data decoding is overall sub-optimal. Therefore, it is desirable to devise a scheme capable of jointly estimating the random permutation and the support of the unknown block-sparse vector in each slot. In addition, it will be interesting to improve the proposed scheme by exploiting the fact that the same set of channels are being estimated across different slots. We also believe that making further use of the large-scale fading coefficients can be a fruitful direction for future research.     

\bibliographystyle{IEEEtran}
\bibliography{IEEEabrv,refrences}
\begin{IEEEbiography}[{\includegraphics[width=1.5in,height=1.25in,clip,keepaspectratio]{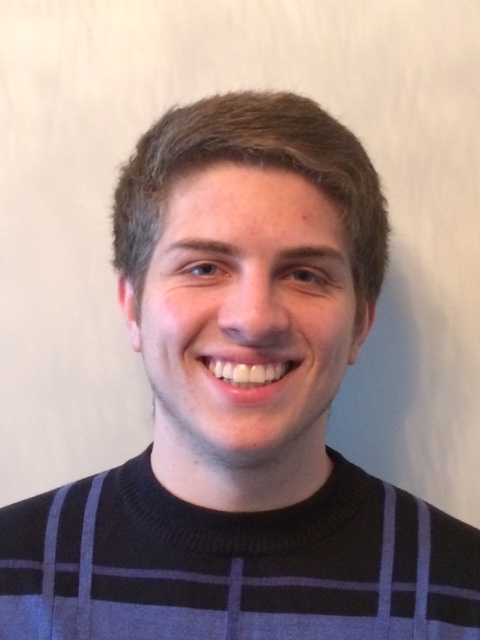}}]%
{Volodymyr Shyianov} is currently working towards completing his B.Sc degree in Electrical Engineering from the University of Manitoba, Canada. He has previously held two Undergraduate Student Research Awards from the Natural Sciences and Engineering Research Council of Canada (NSERC) and one Undergraduate Research Award from the University of Manitoba. His current research interests include Radio-frequency (RF) engineering, multi-user information theory and information theoretically consistent antenna design. 
\end{IEEEbiography}
\begin{IEEEbiography}[{\includegraphics[width=1.5in,height=1.25in,clip,keepaspectratio]{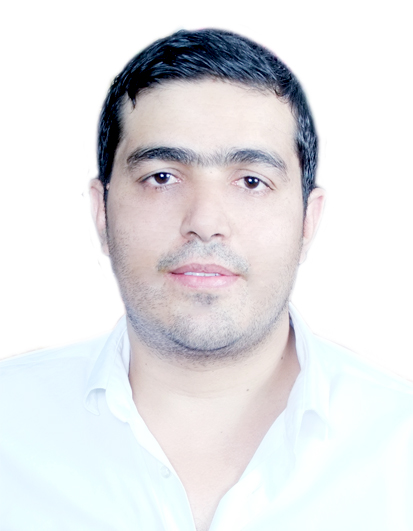}}]%
{Faouzi Bellili} (M'19) received the Diplome d'Ing\'enieur from Tunisia Polytechnic School, in 2007, the M.Sc. and Ph.D. degrees (both with the highest honors) from the Institut national de la recherche scientifique (INRS), University of Quebec, Montreal, QC, Canada, in 2009 and 2014, respectively. He is currently an Assistant Professor with the Department of Electrical and Computer Engineering at the University of Manitoba, Winnipeg, MB, Canada. From Dec. 2016 to May 2018, he was a Postdoctoral Fellow with the ECE Department at the University of Toronto, Toronto, ON,
Canada. From Sept. 2014 to Sept. 2016, he was working as a Research Associate with INRS-EMT where he coordinated a major multi-institutional NSERC Collaborative R\&D (CRD) project on 5th--Generation (5G) Wireless Access Virtualization Enabling Schemes (5G--WAVES). His research focuses on statistical and array signal processing for wireless communications. Dr. Bellili was awarded the very prestigious NSERC PDF Grant over the period 2017-2018. He was also awarded another prestigious PDF Scholarship offered over the same period (but declined) from the ``Fonds de Recherche du Quebec Nature et Technologies'' (FRQNT). He  received the INRS Innovation Award for the year 2014/2015, the very prestigious Academic Gold Medal of the Governor General of Canada for the year 2009-2010, and the Excellence Grant of the Director General of INRS for the year 2009-2010. He received the Award of the best M.Sc. Thesis in INRS-EMT for the year 2009-2010 and twice – for both the MSc and PhD programs – the National Grant of Excellence from the Tunisian Government. In 2011, he was also awarded the Merit Scholarship for Foreign Students from the Ministere de l'Education, du Loisir et du Sport (MELS) of Quebec, Canada. 
\end{IEEEbiography}
\begin{IEEEbiography}[{\includegraphics[width=1.5in,height=1.25in,clip,keepaspectratio]{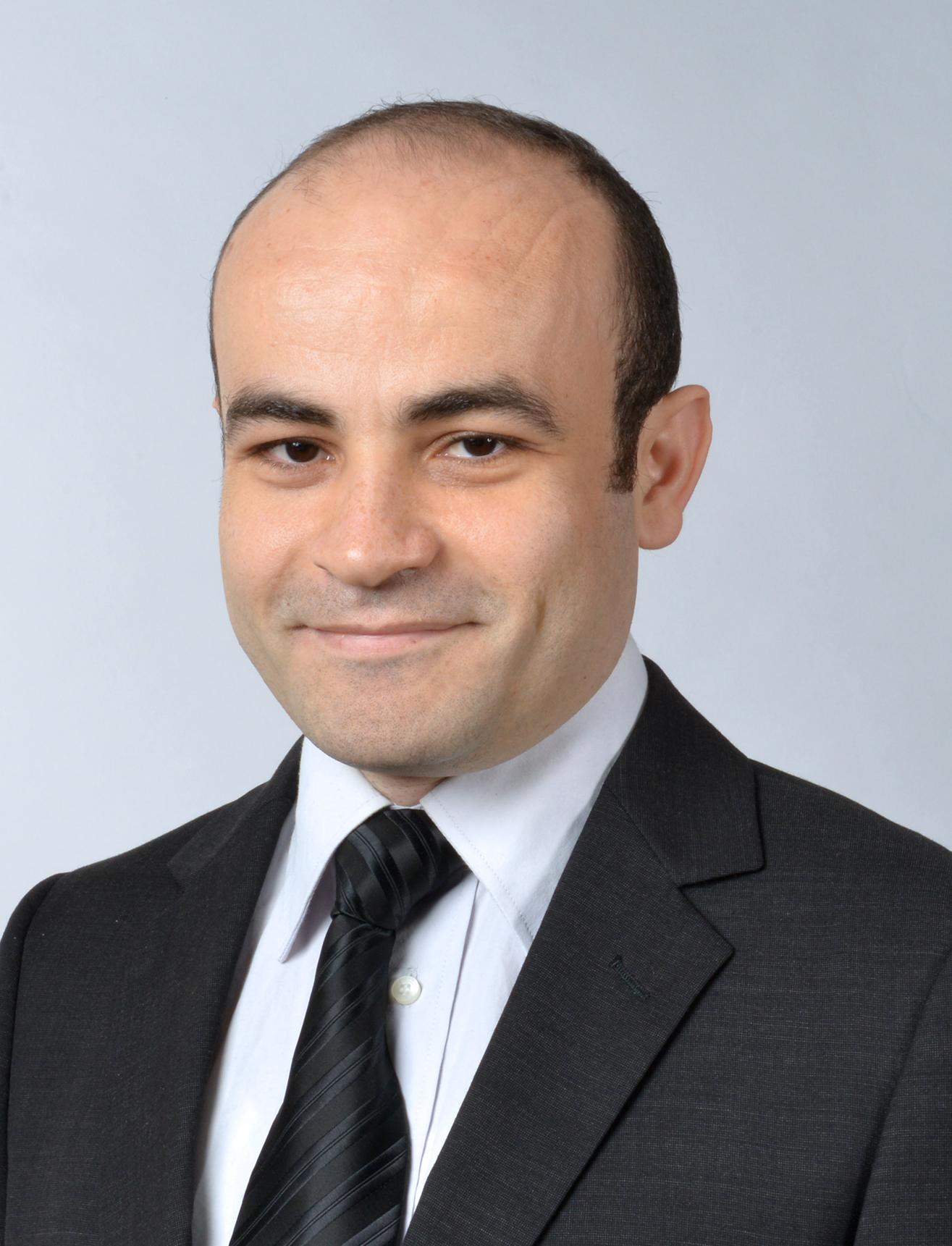}}]%
{Amine Mezghani} (S'08, M'16) received the Ph.D. degree in Electrical Engineering from the Technical University of Munich, Germany in 2015. He is currently an Assistant Professor in the Department of Electrical and Computer Engineering at the University of Manitoba, Canada. Prior to this, he was a Postdoctoral Fellow at the University of Texas at Austin, USA, and a Postdoctoral Scholar with the Department of Electrical Engineering and Computer Science, University of California, Irvine, USA. His current research interests include millimeter-wave communications, massive MIMO, hardware constrained radar and communication systems, antenna theory and large-scale signal processing algorithms. He was the recipient of the joint Rohde \& Schwarz and EE department Outstanding Dissertation Award in 2016.  He has published about hundred papers, particularly on the topic of signal processing and communications with low resolution analog-to-digital and digital-to-analog converters.
\end{IEEEbiography}
 \begin{IEEEbiography}[{\includegraphics[width=1.5in,height=1.25in,clip,keepaspectratio]{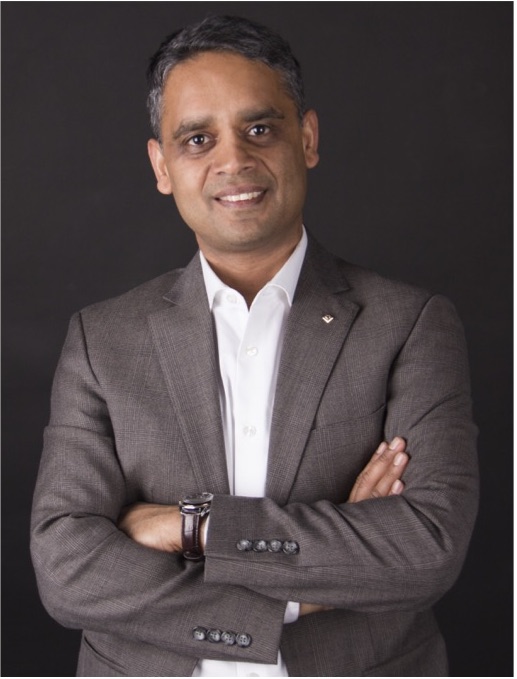}}]%
{Ekram Hossain}(F'15) is a Professor and the Associate Head (Graduate Studies) in the Department of Electrical and Computer Engineering at University of Manitoba, Canada. He is a Member (Class of 2016) of the College of the Royal Society of Canada. Also, he is a Fellow of the Canadian Academy of Engineering. Dr. Hossain's current research interests include design, analysis, and optimization of modern cellular wireless networks, machine learning for wireless communications, and applied game theory. He was elevated to an IEEE Fellow ``for contributions to spectrum management and resource allocation in cognitive and cellular radio networks". He received the 2017 IEEE ComSoc TCGCC (Technical Committee on Green Communications \& Computing) Distinguished Technical Achievement Recognition Award ``for outstanding technical leadership and achievement in green wireless communications and networking”. Dr. Hossain has won several research awards including the “2017 IEEE Communications Society Best Survey Paper Award and the 2011 IEEE Communications Society Fred Ellersick Prize Paper Award. He was listed as a Clarivate Analytics Highly Cited Researcher in Computer Science in 2017, 2018, and 2019. Currently he serves as the Editor-in-Chief of IEEE Press and an Editor for the IEEE Transactions on Mobile Computing. Previously, he served as the Editor-in-Chief for the IEEE Communications Surveys and Tutorials (2012-2016). He is a Distinguished Lecturer of the IEEE Communications Society and the IEEE Vehicular Technology Society. Also, he is an elected member of the Board of Governors of the IEEE Communications Society for the term 2018-2020. 
\end{IEEEbiography}
\end{document}